%% file: main.tex
\documentclass[pdflatex,sn-mathphys-num]{sn-jnl}
\usepackage{graphicx}
\usepackage{multirow}
\usepackage{amsmath,amssymb,amsfonts}
\usepackage{amsthm}
\usepackage{mathrsfs}
\usepackage[title]{appendix}
\usepackage{xcolor}
\usepackage{pifont}
\usepackage{textcomp}
\usepackage{manyfoot}
\usepackage{booktabs}
\usepackage{algorithm}
\usepackage{algorithmicx}
\usepackage{tabularx}
\usepackage{algpseudocode}
\usepackage{caption} 
\usepackage{listings}

\usepackage{subcaption}

\newtheorem{theorem}{Theorem}

\newtheorem{remark}{Remark}

\newtheorem{definition}{Definition}

\begin{document}

\title{The Impact of Egg Quiescence on the Efficacy of \textit{Wolbachia}-Infected Mosquito Releases for Arbovirus Control}

\author*[1]{\fnm{Lu\'is E. S.} \sur{Lopes}}
\email{luiseduardo@alumni.usp.br}
\author[2]{\fnm{Cl\'audia P.} \sur{Ferreira}}
\email{claudia.pio@unesp.br}

\affil[1]{\orgdiv{Department of Applied Mathematics}, \orgname{University of Sao Paulo (USP), Institute of Mathematics and Statistics}, \orgaddress{\city{S\~ao Paulo}, \postcode{05508-090}, \state{S\~ao Paulo}, \country{Brazil}}}

\affil[2]{\orgdiv{Department of Biodiversity and Biostatistics}, \orgname{S\~ao Paulo State University (UNESP), Institute of Biosciences}, \orgaddress{\city{Botucatu}, \postcode{18618-689}, \state{S\~ao Paulo}, \country{Brazil}}}

\abstract{An ordinary differential model is proposed to understand the role of egg quiescence on the efficacy of releasing \textit{Wolbachia}-infected mosquitoes to control arbovirus transmission. The model admits up to five equilibrium points and four biologically meaningful scenarios: extinction of both populations; persistence of the uninfected population with extinction of the infected one; persistence of the infected population with extinction of the uninfected one; and coexistence of both populations. This occurs because the coexistence scenario allows for bistability in the system. A sensitivity analysis shows that mosquitoes optimize their fitness by adjusting the quiescence rate. Because \textit{Wolbachia}-infected eggs do not survive quiescence, or the adults that emerge are infertile, quiescence negatively impacts the fitness of infected mosquitoes, thereby reducing the prevalence of infection in the mosquito population. This increases the chance of encountering an uninfected mosquito, which is more likely to carry the dengue virus compared to a \textit{Wolbachia}-infected one, potentially increasing the risk of dengue transmission during or after environmental stress that triggers quiescence. More importantly, quiescence can compromise the establishment of infection within the mosquito population by increasing the number of infected mosquitoes required during the release period. Therefore, the use of \textit{Wolbachia}-infected mosquitoes to control arbovirus transmission in regions where quiescence occurs at a high rate can be jeopardized.}
\keywords{mathematical model, ordinary differential equation, stability analysis, \textit{Wolbachia}-infection prevalence, risk of dengue transmission}

\maketitle

\section{Introduction}
For several arboviral infections, controlling their vectors remains the most widely used or exclusive method to mitigate pathogen transmission. This approach necessitates an in-depth understanding of the vector's life cycle and its interactions with the environment and other species. Furthermore, it encompasses various strategies that target different stages of the vector's life cycle, which, when combined, can optimally reduce the vector population \citep{yang2008assessing,rafikov2015optimization}. The mosquitoes \textit{Aedes aegypti} and \textit{Aedes albopictus} are the main vectors of Yellow Fever, Dengue, Chikungunya, and Zika viruses in urban settings. Despite decades of efforts to reduce \textit{Aedes} populations, these four diseases remain endemic and epidemic in many countries \citep{kraemer2019past}. Many factors contribute to the emergence, reemergence, and spread of these four diseases. In the absence of an effective and affordable vaccine, newly designed control strategies for \textit{Aedes} mosquitoes are urgently needed (see \href{https://iris.paho.org/handle/10665.2/51375}{Pan American Health Organization}).

In this context, the release of \textit{Wolbachia}-infected mosquitoes has emerged as an alternative strategy to reduce or eliminate the \textit{Aedes aegypti} population — the primary urban vector in tropical regions, and consequently stop the transmission of arboviruses to human populations \citep{pinto2021effectiveness}. The antiviral effect induced by the presence of the symbiont in its host, along with infection traits such as cytoplasmic incompatibility and maternal inheritance, depends on both bacterial density and strain \citep{martinez2014symbionts,axford2016fitness}. 
Therefore, different long-term outcomes are expected — such as population replacement or suppression — following a successful release of either \textit{Wolbachia}-infected eggs or adults \citep{ross2021designing,ross2022decade}. Recently, several experimental and theoretical studies have addressed the sensitivity of bacterial strains to temperature and how this impacts the efficacy of environmentally friendly strategies, such as weakening the reproductive effects induced by the bacterial symbiont \citep{ross2019loss,ross2020heatwaves,mancini2021high,
 ross2023developing,lopes2023exploring}.
 Temperature also influences the life cycle of mosquitoes, affecting fertilization, development, survival, and dormancy \citep{yang2009assessing,
 reinhold2018effects}. 
 
Diapause and quiescence are dormancy mechanisms observed in many species of mosquitoes that allow survival during and after environmental stress. Depending on the species, dormancy can occur at various stages of the mosquito life cycle and is triggered primarily by photoperiod, temperature, and humidity \citep{diniz2017diapause}. Although quiescence affects egg hatch rates and is common in tropical mosquitoes, diapause can impact the larval and adult stages of mosquitoes, particularly in temperate regions. In summary, diapause is a programmed state of developmental arrest influenced by genetic factors. It is seasonal, synchronizes life stages and optimizes population growth, survival, and reproduction
\citep{batz2020rapid}.
On the other hand, quiescence does not depend on endogenous control; it is a non-seasonal, direct, rapid event that is also immediately reversible \citep{oliva2018quiescence}. Quiescent eggs can remain viable for up to 120 days after laying. Coupled with the potential for transovarial transmission of flaviviruses, this increases the vector's capacity to maintain and disseminate pathogens, thereby influencing the dynamics and control of several diseases \citep{ferreira2020silent}. Variations in egg quiescence duration and hatching rates contribute to the persistence and fluctuations of mosquito populations in tropical regions \citep{yang2014assessing}.

Although laboratory, small cage, and semi-field experiments continue to explore the vast diversity of \textit{Wolbachia} strains found in nature, only three have been used in field releases: \textit{w}Mel, \textit{w}MelPop, and \textit{w}AlbB \citep{hoffmann2015wolbachia, ritchie2015application,ant2018wolbachia,dos2022estimating}. In summary, \textit{w}Mel has a lower impact on host fitness compared to the other strains, while \textit{w}AlbB exhibits greater resilience to high temperatures. {\it w}MelPop offers enhanced pathogen blocking capabilities. Other phenotypic traits are similar among these strains  \citep{ferreira2020aedes,
ross2021designing}. Long-term storage of \textit{Aedes aegypti} eggs infected with \textit{Wolbachia} strains — specifically \textit{w}Mel, \textit{w}AlbB, and \textit{w}MelPop — can significantly reduce egg viability. In addition, adult females emerging from these stored eggs often exhibit infertility \citep{mcmeniman2010virulent,lau2021infertility,ross2020persistent}.
In response, new lines and variants — \textit{w}Au, \textit{w}AlbA, \textit{w}MelM —  have been characterized \citep{ant2018wolbachia,gu2022wmel}.

Although the direct effects of temperature on the reduction of cytoplasmic incompatibility and maternal inheritance have been addressed by mathematical models, stressful environmental conditions that trigger dormancy mechanisms have not. Here, focusing on quiescence — a trait characteristic of \textit{Ae. aegypti} mosquitoes — we investigate its impact on fitness of both uninfected and \textit{Wolbachia}-infected mosquitoes, as well as on prevalence of infection, using a mathematical model.
Quiescent eggs and their role in disease transmission have been explored in the literature \citep{erguler2016large,yang2017transovarial,pliego2017seasonality}, but their relationship with the release of \textit{Wolbachia}-carrying mosquitoes has not been addressed yet.

\section{Mathematical Model}

The uninfected and \textit{Wolbachia}-infected mosquito populations are divided into seven compartments: eggs, larvae plus pupae, adults, and quiescent (latent) eggs, denoted by $O$, $O^w$, $I$, $I^w$,  $A$, $A^w$, and $Q$, respectively. The superscript $w$ indicates the \textit{Wolbachia}-infected population. The $Q^w$ compartment is excluded from the model, as infected eggs exhibit higher mortality rates compared to uninfected ones, and the adult females emerging from them are infertile \citep{mcmeniman2010virulent,lau2021infertility,ross2020persistent}.

The sex ratio and {\it per-capita} oviposition rate are denoted by $r$ and $r^w$, and $\phi$ and $\phi^w$, respectively. Consequently, $\phi rA$  and $\phi (1-r)A$ represent the proportions of females and males emerging from an uninfected female mosquito per unit time. Similarly,  $\phi^w r^w A^w$ and $\phi^w (1-r^w) A^w$ represent the proportions of females and males emerging from an infected female mosquito per unit time. Therefore, the ratios
\[\frac{(1-r)A}{(1-r)A+(1-r^w)A^w} \quad \mbox{and} \quad  \frac{(1-r^w)A^w}{(1-r)A+(1-r^w)A^w}\]
can be interpreted as the probabilities of mating with uninfected and infected males, respectively. However, not all matings between uninfected females and infected males result in fertile eggs. The success of these matings depends on cytoplasmic incompatibility (CI), denoted by $\nu$ $\in [0,1]$. If $\nu=1$, all eggs generated from this mating are fertile, while if $\nu=0$ all of them are infertile. Therefore, 
\[
\phi \nu rA  \frac{(1-r^w)A^w}{(1-r)A+(1-r^w)A^w}
\]
represents the proportion of fertile eggs produced per unit time from a successful mating between an uninfected female and an infected male.

Vertical transmission of \textit{Wolbachia} depends on maternal inheritance, denoted by $\zeta$  $\in [0,1]$. If $\zeta=1$, all eggs from an infected female mosquito will also be infected; whereas when $\zeta=0$, none of the eggs are infected. Therefore,
\[(1-\zeta) \phi^w r^w A^w \quad \mbox{and} \quad \zeta \phi^w r^w A^w\] represent the proportions of uninfected and infected eggs laid per unit time by an infected female, respectively. The parameters $\sigma_o$, $\sigma^w_o$, $\eta_o$, $\eta^w_o$, $\mu_o$, and $\mu^w_o$ represent the {\it per-capita} rates of egg hatching,  quiescence, and natural mortality for uninfected and \textit{Wolbachia}-infected mosquitoes, respectively, with the infected parameters denoted by the superscript $w$.

The two expressions
\[(\sigma_o O + \sigma_q Q) \left(1-\frac{I+I^w}{k}\right) \quad \mbox{and} \quad \sigma^w_o O  \left(1-\frac{I+I^w}{k}\right)\]
represent the recruitment rates of uninfected and infected immature stages (larvae plus pupae populations), respectively, considering the density-dependent intraspecific competition occurring primarily during the larval stage. The parameters $\sigma_q$ and $\mu_q$ denote the {\it per-capita} rates of hatching from the quiescent stage into larvae and natural mortality during the quiescent stage, respectively. The parameters $\sigma_i$, $\sigma^w_i$, $\mu_i$, $\mu^w_i$ represent the \textit{per-capita} rates of immature development and natural mortality during the immature stage for uninfected and \textit{ Wolbachia}-infected mosquitoes, respectively. Lastly, $\mu_a$ and $\mu^w_a$ represent the {\it per-capita} natural mortality rates of adult \textit{Wolbachia}-infected and uninfected mosquitoes, respectively.

Therefore, the proposed nonlinear ordinary differential model is given by
\begin{eqnarray}
&&\frac{dO}{dt}= \phi rA \left[\frac{(1-r)A+ \nu(1-r^w)A^w}{(1-r)A+(1-r^w)A^w}\right] +  \phi^w r^wA^w(1-\zeta)-
 O(\sigma_o + \eta_o + \mu_o) \nonumber\\
&&\frac{dI}{dt}=(\sigma_o O + \sigma_q Q) \left(1- \frac{I+I^w}{k}\right) - I(\sigma_i + \mu_ i) \nonumber \\
&&\frac{dA}{dt}= \sigma_i I-\mu_a A \label{systemO}\\
&&\frac{dQ}{dt}= \eta_o O- Q(\sigma_q + \mu_q) \nonumber\\
&&\frac{dO^w}{dt}= \phi^w \zeta r^wA^w - O^w(\eta_o^w+ \mu_o^w+\sigma_o^w) \nonumber \\
&&\frac{dI^w}{dt}= \sigma_o^w O^w \left(1- \frac{I+I^w}{k}\right)- I^w(\sigma_i^w + \mu_i^w) \nonumber \\
&&\frac{dA^w}{dt}= \sigma_i^w I^w- \mu_a^wA^w. \nonumber 
 \end{eqnarray}
In system \eqref{systemO},  the variables $O:=O(t)$, $O^w:=O^w(t)$, $I:=I(t)$, $I^w:=I^w(t)$,  $A:=A(t)$, $A^w:=A^w(t)$, $Q:=Q(t)$, represent the populations of uninfected eggs, infected eggs, uninfected immature, infected immature, uninfected adults, infected adults, and quiescent eggs, respectively, as functions of time $t$ with $t\in[0,+\infty)$. Furthermore, all parameters and variables in the model are positive. In particular, the carrying capacity $k$ is strictly positive. Table \ref{tab:parametros} lists the parameters of the model with their biological interpretations and units. 
 
\begin{table}[ht]
    \caption{Model's parameters, their biological interpretation, and units. The subscript $j=\{o,i,a,q\}$ designates non-quiescent egg, immature, adult, and quiescent egg. The superscript $w$ indicates that the respective entomological parameter belongs to the \textit{Wolbachia}-infected mosquito.\vspace{0.2cm}}

      \begin{tabular}{llc} \hline
         Parameters  & Biological Interpretation & Units\\\hline
         $\phi,\phi^w$  & 
         {\it Per-capita} oviposition rates &[time]$^{-1}$ \\
         $\nu$ & 
         Cytoplasmic incompatibility &-\\
         $\zeta$ & Maternal inheritance &- \\
         $k$ & Carrying capacity &[individuals]\\
         $\sigma_j,\sigma_j^w$,$\eta_o,\eta_o^w$ & {\it Per-capita} transition rates  & [time]$^{-1}$\\
         $\mu_j,\mu_j^w$ & {\it Per-capita} mortality rates &[time]$^{-1}$\\
         $r,r^w$ & Sex ratio&- \\
        \hline
      \end{tabular}
      \label{tab:parametros}
  \end{table}

\section{Results}
\subsection{Existence, Positivity, and Boundedness of Solutions}
The following results establish the existence, positivity, and boundedness of the solutions to system \eqref{systemO} for all  $t\in [0, +\infty)$. To facilitate this, let us define $X(t):=\left(O(t),~ I(t),~ A(t),~ Q(t),~ O^w(t),~ I^w(t),~A^w(t)\right)$ as the state vector of the system \eqref{systemO} at any time $t$, given an initial condition $X(0)$. In addition, $N(t):=O(t)+I(t)+A(t)+Q(t)$ and $N^w(t):=O^w(t)+I^w(t)+A^w(t)$ denote the total populations of uninfected and \textit{Wolbachia}-infected mosquitoes, respectively.

\begin{theorem} \label{theorem1} If $X(0)>0 \mbox{ and } (I+I^w)(0)<k$,  then, for all $t>0$,  $X(t)$ exits, it is unique, and satisfies 
\begin{eqnarray}
X(t)>0 \mbox{ and } (I+I^w)(t)<k. 
\end{eqnarray}
\end{theorem}

\begin{theorem} \label{theorem2} Assume that $X(0)>0 \mbox{ and } (I+I^w)(0)<k$.  Then, there are constants $\tau>0$ and $\tau_w>0$ such that, for all $t>0$, we have  $N(t) \leq \tau$ and $N^w(t) \leq \tau_w.$ 
\end{theorem}

The proofs of Theorems 1 and 2 are done in Appendix \ref{proof1}.

\subsection{Existence and Stability of the Equilibrium Points}\label{stabilityAequilibrium}

In this section, we analyze the existence and stability of the equilibrium points of system \eqref{systemO}. Let us start defining $\underline r:= 1-r$, $\underline r^w:=1-r^w$, $\underline \zeta:=1-\zeta$,
$\lambda_o:=\sigma_o + \eta_o + \mu_o$, $\lambda_o^w:=\sigma_o^w + \eta_o^w + \mu_o^w$, $\lambda_i:=\sigma_i + \mu_i$, $\lambda_i^w=\sigma_i^w + \mu_i^w$,  $\lambda_q:=\sigma_q + \mu_q, b:=\sigma_i/\mu_a, b^w:=\sigma_i^w/\mu_a^w, c:=\sigma_o + (\sigma_q \eta_o)/\lambda_q,$ and $d^w:= \phi^w \zeta r^w b^w
$. Thus, by solving the nonlinear system with the derivatives set to zero, we can obtain the equilibrium solutions (refer to Appendix~\ref{existence_equilibrium}) and the thresholds for their existence. Subsequently, the local asymptotic stability of each equilibrium can be analyzed using various techniques (see Appendix~\ref{satibility_conditions}).
\begin{theorem}
    \label{steady_states} System \eqref{systemO} has up to five equilibrium points:

\begin{enumerate}
    \item [(i)] 
$P_{(0,0)}=(0,0,0,0,0,0,0)$ which always exists;

    \item [(ii)] 
$P_{(u,0)}=\left(\dfrac{k\lambda_i (R_u-1)}{c}, \dfrac{k(R_u-1)}{R_u}, \dfrac{bk(R_u-1)}{R_u}, \dfrac{\eta_ok\lambda_i(R_u-1)}{\lambda_qc},0,0,0\right)$ that exists if $R_u>1$ with $\displaystyle{R_u:= \frac{\phi r b  c}{ \lambda_o\lambda_i}}$;

    \item [(iii)]  
$P_{(0,w)}=\left(0,0,0,0,\dfrac{k\lambda_i^w\left(R_w-1\right)}{\sigma_o^w}, \dfrac{k\left(R_w-1\right)}{R_w}, \dfrac{b^w k\left(R_w-1\right)}{R_w}\right)$ that exists if $\zeta=1$ and  $R_w>1$, with $\displaystyle{R_w:=\frac{d^w \sigma_o^w}{\lambda_i^w\lambda_o^w}}$;  

    \item [(iv)] $P_{(u,w)}=(\bar O,\bar I,\bar A,\bar Q,\bar O^w, \bar I^w, \bar A^w)$ that exists if $\zeta \neq 1,~ R_w >1,$ and
     $\bar O$ satisfies 

\[A_1\bar O^2+B_1\bar O+C_1=0\]where
\begin{eqnarray}
    A_1 &=& -c [\underline r b (R_w+R_{uw}-R_u)-\underline r^w b^w (R_w+R_{uw}-\nu R_u)],\nonumber\\
    B_1 &=& k \lambda_i(R_w -1) [\underline r b R_{uw}-\underline r^w b^w(R_w +2R_{uw}-\nu R_u)], \nonumber \\
    C_1 &=&\frac{(k\lambda_i)^2(R_w- 1)^2}{c}\underline r^wb^wR_{uw}>0; \quad R_{uw}:=\frac{\phi^w r^wb^w\underline \zeta c}{\lambda_o\lambda_i},\nonumber
\end{eqnarray} and \[0<\bar O< \frac{k\lambda_i(R_w-1)}{c}.\]

Descartes' rule of signs and the discriminant's positivity conditions help determine if the polynomial admits one or two positive real roots. Therefore:

\begin{enumerate}
    \item a unique $P_{(u,w)}$ exists if one of the following conditions holds:
        \begin{enumerate}
            \item [a1.] $\nu=1$ and $R_w > R_u$.
            
        \item [a2.] $\nu \in [0,1)$, $R_{\Delta}=1$ and $S_2 \in (2S_1,0) \cup(0,2S_1),$ where \begin{eqnarray}
    R_{\Delta} &:=& \frac{ \left[ \underline{r} b R_{uw} + \underline{r}^w b^w (R_w - \nu R_u) \right]^2 }{ 4 \underline{r} b \underline{r}^w b^w R_{uw} R_u (1 - \nu) },\nonumber\\    
    S_1 &:=&\underline r b (R_w+R_{uw}-R_u)-\underline r^w b^w (R_w+R_{uw}-\nu R_u),\nonumber\\
    S_2&:=& \underline r b R_{uw}-\underline r^w b^w(R_w +2R_{uw}-\nu R_u).\nonumber
\end{eqnarray}

        \item [a3.] $\nu \in [0,1)$, $\underline r b \neq \underline r^w b^w$, $S_1>0$ and $S_2 \in (-S_3,\, 2S_1 - S_3)$, where \begin{equation*}
    S_{3}:=2\sqrt{\underline r b \underline r^w b^w R_{uw} R_u (1-\nu)(R_{\Delta}-1)}>0.
\end{equation*}
        \end{enumerate}
    
    \item two solutions \( P_{(u,w)}^- \) and \( P_{(u,w)}^+ \) exist if the following conditions are satisfied:
\[
\nu \in [0,1),~ R_{\Delta} > 1,~ S_1 < 0,~ S_2 < 0,~ \text{and}~ S_2  \in (2S_1-S_3, -S_3)\cap(2S_1+S_3,0).
\]
\end{enumerate}

\end{enumerate}
\end{theorem}

The explicit expressions for the coexistence equilibrium values under the above conditions are provided in Appendix~\ref{existence_equilibrium}.

Furthermore,  $P_{(0,0)}$ is called the trivial equilibrium and represents the extinction of both populations;  $P_{(u,0)}$ is the infection-free equilibrium, where the uninfected population persists and the infected population goes extinct;  $P_{(0,w)}$ represents the extinction of the uninfected population and the persistence of the infected one; and $P_{(u,w)}$ is the coexistence equilibrium at which both uninfected and infected populations persist.

\begin{theorem} Consider the system \eqref{systemO}. The local stability of the equilibrium points is given by:
\begin{itemize}
    \item [(i)]  If $\max\{R_u,  R_w \}<1$, then $P_{(0,0)}$ is locally asymptotically stable;
    \item [(ii)]If $ R_u>\max\{1,  R_w \}$,  then $P_{(u,0)}$ is locally asymptotically stable;
    \item [(iii)]  If $ R_w>\max\{1, \nu  R_u \}$ and $\zeta=1$, then $P_{(0,w)}$ is locally asymptotically stable.
\end{itemize}

\end{theorem}
These two thresholds, $R_u$ and $R_w$, have biological interpretations.  The first measures the fitness of the uninfected population, while the second quantifies the fitness of the \textit{ Wolbachia}-infected population when they are isolated, that is, when they do not share the same geographical space. On the other hand, the parameter $R_{uw}$, which appears in the coefficient of the polynomial that determines when coexistence appears, measures the contribution of \textit{Wolbachia}-infected mosquitoes — that produce uninfected offspring due to imperfect vertical transmission ($\zeta \neq 1$) — to the fitness of the uninfected population. Thus, the value of $R_{uw}$ plays an important role in determining the relative fitness of the two populations during competition. 

\subsection{Quiescence}
\label{sensitivity_quiescence}

A simple way to measure the effect of the quiescence rate on the fitness of the uninfected and \textit{Wolbachia}-infected populations is through the normalized sensitivity index.

\begin{definition}
The normalized sensitivity index of a variable $f$ to a parameter $g$ is defined as \cite{ChitnisSensitivity} 
\[\Gamma_g^f:=\frac{g}{f}\frac{\partial f}{\partial g}.\]
\end{definition}

Thus, the sensitivity index of $R_u$ to $\eta_o$ is
 \[\Gamma_{\eta_o}^{R_u}=\frac{\eta_o}{R_u}\frac{\partial R_u}{\partial \eta_o} =\frac{\eta_o \left(\mu_o\sigma_q-\sigma_o\mu_q\right)}{\lambda_o(\sigma_o\lambda_q+\sigma_q\eta_o)}\Longrightarrow
 \Delta R_u \approx \frac{R_u \left(\mu_o\sigma_q-\sigma_o\mu_q\right)}{\lambda_o(\sigma_o\lambda_q+\sigma_q\eta_o)} \Delta \eta_o.
 \]

 Keeping all parameters fixed except for $\eta_0$, two situations are observed:
 \begin{enumerate}
 \item[(i)] when $\mu_o\sigma_q-\sigma_o\mu_q>0$, the egg ceases development and remains in a quiescent state. This implies that $\Delta R_u>0$ since the rate of quiescence increases.
 \item[(ii)]  when $\mu_o\sigma_q-\sigma_o\mu_q<0$,
 the egg continues its development and hatches as a larva. This implies that $\Delta R_u>0$ since the rate of quiescence decreases. 
 \end{enumerate}

The first scenario describes a situation where the abiotic conditions are not good for the mosquito population (for example, low and high temperatures or low humidity), while the second corresponds to a situation where the abiotic conditions are good \citep{yang2009assessing,farnesi2009embryonic,eisen2014impact,martin2025desiccation}. Therefore, in response to abiotic cues, the uninfected mosquito optimizes its fitness by increasing either the hatching rate or the quiescence rate of its eggs. Interestingly, $\Gamma_{\eta_o}^{R_u} = \Gamma_{\eta_o}^{R_{uw}}$, indicating that although $ R_u $ and $ R_{uw} $ represent distinct fitness components, the environmental or behavioral regulation mediated by quiescence exerts an equivalent proportional effect on both.

Moreover, the sensitivity index of $R_w$ with respect to $\eta_o^w$ is given by
\[
\Gamma_{\eta_o^w}^{R_w} = \frac{\eta_o^w}{R_w} \frac{\partial R_w}{\partial \eta_o^w} = -\frac{\eta_o^w}{\lambda_o^w} \Longrightarrow \Delta R_w \approx -\frac{R_w}{\lambda_o^w} \Delta \eta_o^w.
\]

Thus, an increase in the rate of quiescence reduces the fitness of the uninfected mosquito and may compromise its long-term persistence. This suggests that quiescent eggs contribute negatively to the efficacy of the \textit{Wolbachia}-infected mosquito release strategy, as quiescence compromises the viability of the eggs, the fertility of the adults \citep{mcmeniman2010virulent, lau2021infertility, ross2020persistent}, and, consequently, the success of the population replacement effort.

\subsection{Numerical Experiments}

The ordinary differential system proposed in \eqref{systemO} was solved using the SciPy library in Python. The baseline parameter set used in the simulations is: $\sigma_i =\sigma_i^w = 0.11$, $\sigma_o = \sigma_o^w = 0.09$, $\mu_i = \mu_{i}^w = 0.08$, $\mu_o = \mu_{o}^w = 0.015$, $\mu_a = \mu_{a}^w = 0.03$, $\sigma_q = 0.002$, $\mu_q = 0.004$, $\phi =0.24$, $\phi^w =0.35$ on days$^{-1}$, $k = 1500$ individuals, $r = r^w = 0.5$, $\zeta=0.8$, and $\nu = 0.1$ (see Table \ref{paraVal}).

\begin{table}[ht]
 \caption{Parameters with their range's values, units, and key references.}\vspace{0.2cm}

    \begin{tabular}{llll} \hline
        Parameters  & Range  & Unit & References  \\\hline
       $\nu$ &  $[0,1]$ & - & \cite{lopes2023exploring, ferreira2020aedes, hoffmann2014stability}  \\
       $\zeta$ &$[0,1]$ &-& \cite{lopes2023exploring, ferreira2020aedes, walker2011wmel} \\
       $r, r^w$ &  $[0,1]$& -& \cite{lopes2023exploring, ferreira2020aedes}\\
       $\phi, \phi_w$&$(0,11.5], (0,9.8]$ 
       & [days]$^{-1}$
& \cite{lopes2023exploring, yang2009assessing, farnesi2009embryonic} \\
       $\sigma_o, \sigma_o^w, \sigma_q$&$(0, 0.11], (0, 0.11], (0,0.03]$&[days]$^{-1}$ &\cite{farnesi2009embryonic, yang2014assessing, silva1999influencia}\\
       $\sigma_i, \sigma_i^w$&$ (0,0.12] $&[days]$^{-1}$& \cite{eisen2014impact, marinho2016effects, yang2014assessing, silva1999influencia}\\
       $\eta_o, \eta_o^w$ & $(0,0.2]$&[days]$^{-1}$&\cite{oliva2018quiescence,silva1999influencia, yang2014assessing}\\
       $\mu_o, \mu_o^w, \mu_q$ &$(0, 0.2], (0,0.3], (0,0.27]$ &[days]$^{-1}$
& \cite{yang2014assessing, silva1999influencia}\\
       $\mu_i, \mu_i^w$ & $(0,0.45], (0,5]$& [days]$^{-1}$& \cite{Sukiato2019,lopes2023exploring,yang2009assessing}\\
       $\mu_a, \mu_a^w$ & [0.02, 0.1], [0.03, 0.11]&[days]$^{-1}$&\cite{lopes2023exploring, yang2009assessing}\\ 
       $k$&$[100,1500]$&individuals& Assumed\\\hline
      \end{tabular}
      \label{paraVal}
  \end{table}

\subsubsection{Existence and Local Stability of the Equilibrium Points}

Let's examine regions of the parameter space where $R_u$ and $R_w$ are either greater or less than 1. In all figures, to vary $R_u$ and $R_w$, we change either $\eta_o$ or $\eta_o^w$. This allows us to confirm the analytical results obtained in Section \ref{stabilityAequilibrium} and to explore the coexistence equilibrium, for which we were unable to analyze the stability analytically.

\begin{figure}
\centering
\begin{subfigure}{.48\textwidth}
  \centering
  \includegraphics[width=1\linewidth]{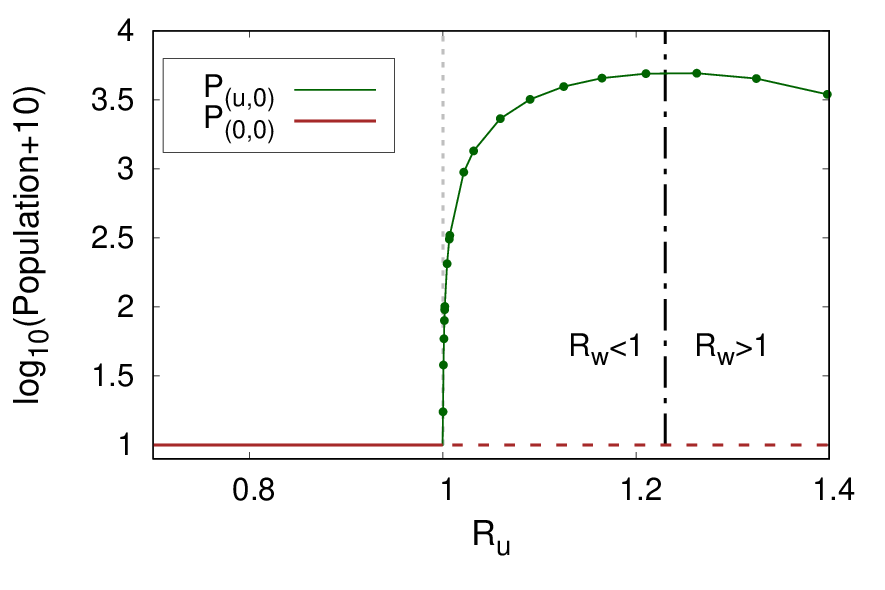}
  \caption{$\phi=\phi^w=0.17,  \zeta=0.98$}
  \label{fig:P_u0_to_P00}
\end{subfigure}
\begin{subfigure}{.48\textwidth}
  \centering
  \includegraphics[width=1\linewidth]{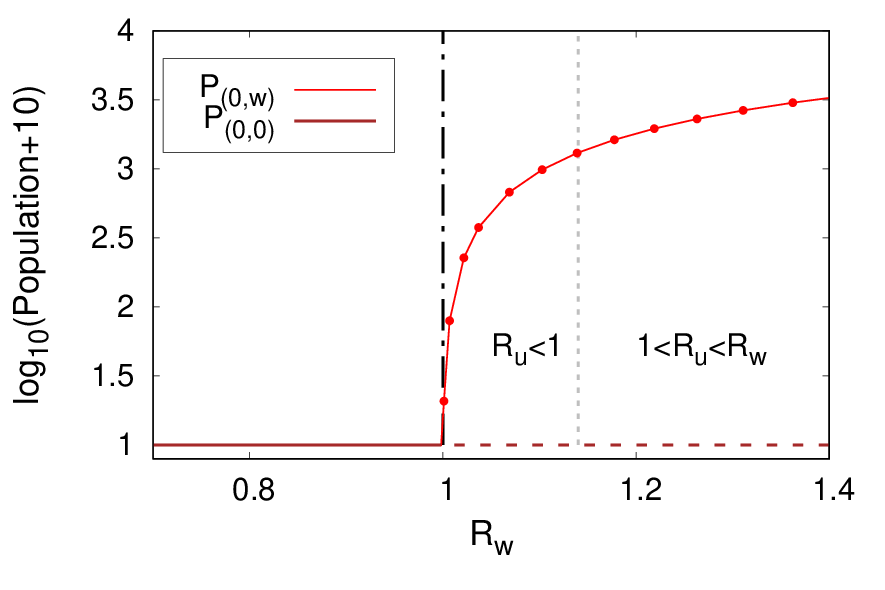}
  \caption{$\phi=0.2, \phi^w=0.4,  \zeta=1$}
  \label{fig:P_00_to_P0w}
\end{subfigure}

\begin{subfigure}{.48\textwidth}
  \centering
 
  \includegraphics[width=1\linewidth]{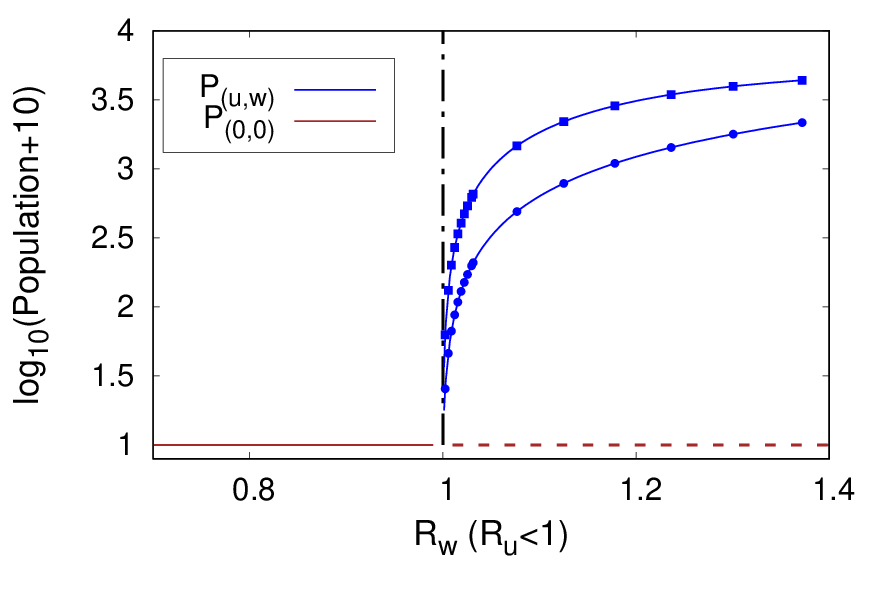}
  \caption{$\phi=0.15, \phi^w=0.36$ and $\zeta=0.8$}
  \label{fig:P_00_to_Puw}
\end{subfigure}
\begin{subfigure}{.48\textwidth}
  \centering
  \includegraphics[width=1\linewidth]{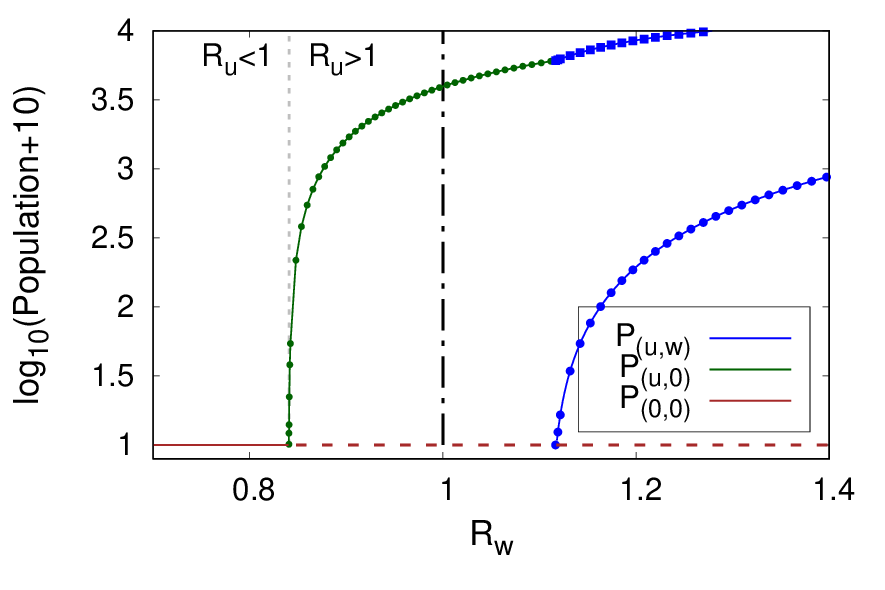}
  \caption{$\phi=0.2, \phi^w=0.36$ and $\zeta=0.8$}
  \label{fig:P_u0_to_Puw}
\end{subfigure}
\begin{subfigure}{.5\textwidth}
  \centering
  \includegraphics[width=1\linewidth]{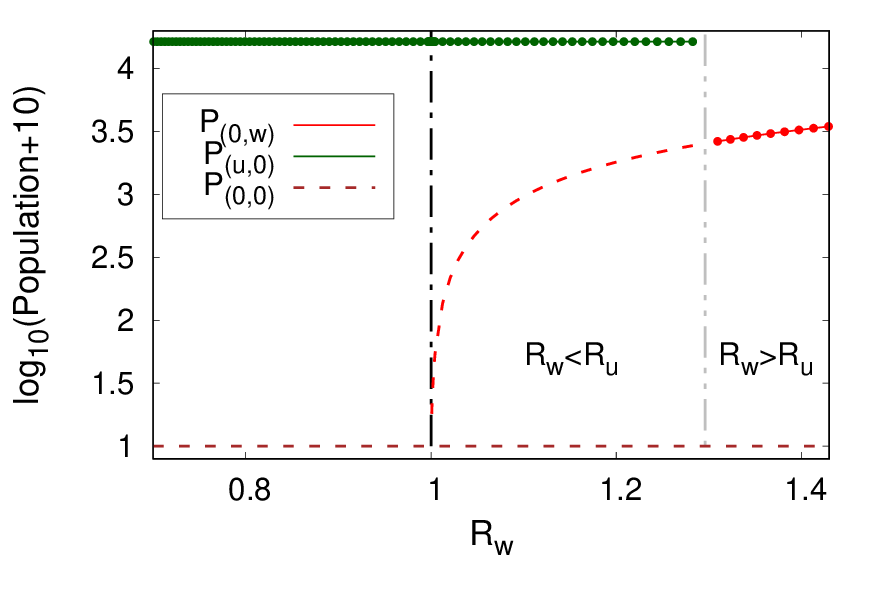}
  \caption{ $\phi=0.24, \phi^w=0.288, \eta_o=0.14$ and $\zeta=1$ }
  \label{fig:P_u0_to_Puw1}
\end{subfigure}
\caption{Bifurcation diagrams. Dashed lines mean that the steady state is unstable, while continuous line means the stability of it. The red, green, and blue symbols are simulation results while the lines are analytical ones. For  $P_{(u,0)}$ and $P_{(0,w)}$, only the components different from zero are displayed. In all cases, we are plotting the total uninfected $N=O+I+Q+A$ and infected populations $N^w=O^w+I^w+A^w$. In the case of the coexistence equilibrium, the square (\textcolor{blue}{\scalebox{0.6}{\ding{110}}}) and circles (\textcolor{blue}{\scalebox{0.6}{\ding{108}}}) symbols represent $N$ and $N^w$, respectively. The vertical lines highlight inequalities related to 
$R_u$ and  $R_w$.}
\label{BifurcationDiagrams}
\end{figure}

In Figure~\ref{BifurcationDiagrams}, we plot the equilibrium points across different values of the threshold $R_u$, except for subfigure \ref{fig:P_u0_to_P00}, where we vary $R_w$. In all cases, we set $\nu=1$. When $R_u>1$, the simulation stars at the equilibrium value $P_{(u,0)}$. At this point, \textit{Wolbachia}-infected adult mosquitoes are introduced at a level $A^w=\bar A/2$ where $\bar A$ is the value of $A$ in $P_{(u,0)}$.  All other components are initialized to zero. When $R_u<1$,  $A^w$ was chosen randomly within the interval of $(0,100)$. The green, red, brown, and blue colors specify
$P_{(u,0)}$, $P_{(0,w)}$, $P_{(0,0)}$, $P_{(u,w)}$, respectively, the persistence of the uninfected population and extinction of the infected one, the persistence of the infected population and extinction of the uninfected one, the extinction of both populations, and the persistence of both populations.
We can see that the existence and stability of $P_{(u,0)}$ are guaranteed by $R_u>~\max\{1,R_w\}$ (subfigure \ref{fig:P_u0_to_P00}), while for $P_{(0,w)}$ existence and stability are given by $R_w>\max\{1,R_u\}$ and $\zeta=1$ (subfigure \ref{fig:P_00_to_P0w}). This implies that the total replacement of the uninfected population by a \textit{Wolbachia}-infected one is only possible when maternal transmission is perfect. In general, the population size gradually increases from zero in scenarios where mosquitoes are introduced into an environment free of mosquitoes that offers conditions suitable for their persistence (subfigures \ref{fig:P_u0_to_P00}, \ref{fig:P_00_to_P0w}, and \ref{fig:P_00_to_Puw}), while abrupt changes in mosquito population size (and type) are observed when partial or complete population replacement occurs (subfigures~\ref{fig:P_u0_to_Puw} and \ref{fig:P_u0_to_Puw1}).

\begin{figure}
\centering
\begin{subfigure}{.45\textwidth}
  \centering  \includegraphics[width=1\linewidth]{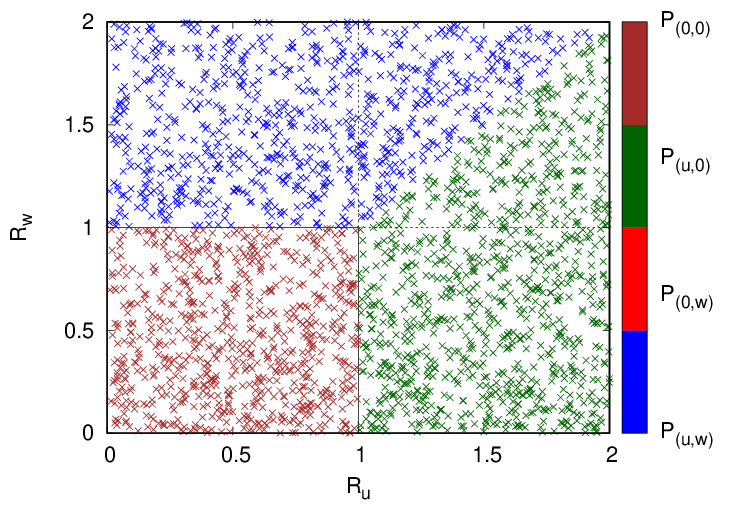}
  \caption{$\zeta \in (0,1)$}
  \label{fig:sub1}
\end{subfigure}
\begin{subfigure}{.45\textwidth}
 \centering
 \includegraphics[width=1\linewidth]{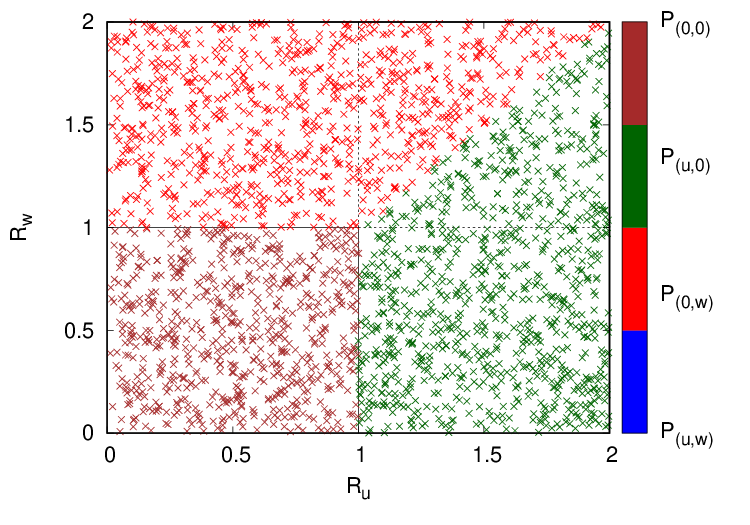}
  \caption{$\zeta =1$}
  \label{fig:sub2}
\end{subfigure}
\caption{$R_u\times R_w$ with blue, green, red, and brown colors showing the steady states  $P_{(u,w)}$, $P_{(u,0)}$, $P_{(0,w)}$ and $P_{(0,0)}$, respectively.}
\label{RuRwcase}
\end{figure}

A more general result is shown in Figure~\ref{RuRwcase}.  Five thousand sets of parameters were randomly sampled from the values displayed in Table \ref{paraVal}, each corresponding to a pair $(R_u,R_w)$. For each parameters set, equilibrium was considered achieved when the least-squares regression line fitted to the uninfected and infected populations ($N$ and $N^w$, respectively) over the last $1.46 \times 10^{5}$ time steps ( corresponding to the last five years of simulation) had a slope less than $10^{-10}$ \citep{caswell1993ecological}. The initial conditions were set to be near (within 10\% above or below) the corresponding equilibrium point obtained analytically and associated with the parameters set. The result is summarized in Table~\ref{Table_rurw}. 

\begin{table}[ht]
\small
\caption{Conditions for existence and stability of equilibrium points. For the coexistence equilibrium
given by $P_{(u,w)}$ or  $P^-_{(u,w)}$ and $P^+_{(u,w)}$, the conditions $\zeta \ne 1$ and $R_w > 1$ must hold.} 
\label{Table_rurw}
\renewcommand{\arraystretch}{1.4}
\begin{tabularx}{\textwidth}{p{2.8cm} X X}
\hline
\textbf{Equilibrium} & \textbf{Existence} & \textbf{Local stability} \\
\hline
$P_{(0,0)}$ & always & $1>\max\left\{R_u, R_w \right\}$ \\
$P_{(u,0)}$ & $R_u > 1$ & $R_u > \max\{1, R_w\}$ \\
$P_{(0,w)}$ & $R_w > 1$ and $\zeta = 1$ & $R_w > \max\{1, \nu R_u\}$ and $\zeta = 1$ \\
\multirow{1}{=}{$P_{(u,w)}$}
& (i) $\nu = 1$,   $R_w > R_u$ or
& \\
& (ii) $\nu \ne 1$, $R_\Delta = 1$,  \\
& $S_2 \in (2S_1,0) \cup (0, 2S_1)$ or
& \\
& (iii) $\nu \ne 1$, $\underline r b \ne \underline r^w b^w$, $S_1 > 0$, $S_2 \in (- S_3, 2S_1-S_3)$ or
& \\
\multirow{1}{=}{$P_{(u,w)}^-$ and $P_{(u,w)}^+$}
& (a)  $\nu \ne 1$, $R_\Delta > 1$, \\
& $S_1 < 0$, $S_2 < 0$,  \\
& $S_2 \mp S_3 \in (2S_1 , 0)$
& \\
\hline
\end{tabularx}
\end{table}

\begin{figure}
\centering
\includegraphics[width=0.6\linewidth]{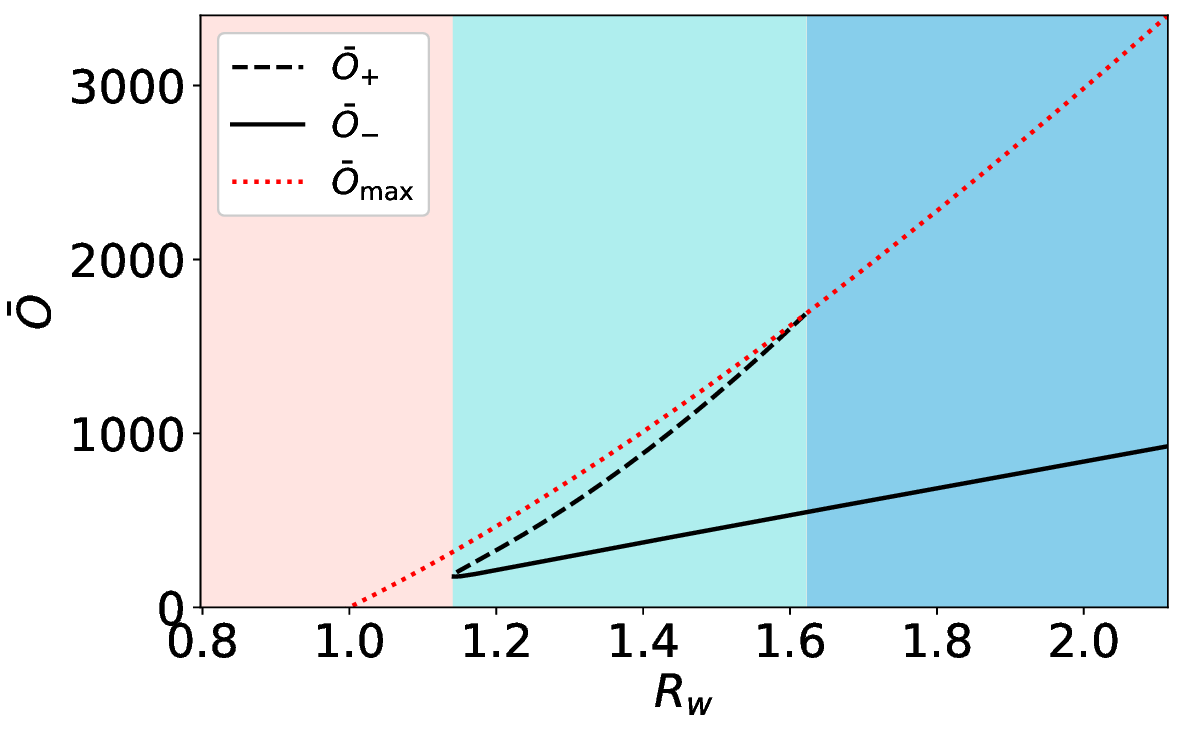}
\caption{Regions of existence of a strictly positive real root of $f(\bar O)=A_1 \bar O^2 + B_1 \bar O + C_1$. Below the red line $0<\bar O < \bar O_{max}$ where $\bar O_{max}:=k\lambda_i(R_w-1)/c$.  The light cyan, blue, and pink regions represent the parameter sets for which the equation admits two, one, or no real roots, respectively. The continuous and dashed black lines represent the stable $\bar O_-$ and unstable equilibria $\bar O_+$ , respectively.} 
\label{fig:existenceO}
\end{figure}

More complex dynamics emerge when both populations are present. In this case, $\zeta \ne 1$ and $R_w>1$. The existence and number of equilibrium points — denoted by \( P_{(u,w)} \), \( P_{(u,w)}^- \) and \( P_{(u,w)}^+ \) — depend on several conditions (see Theorem \ref{steady_states}). For example, coexistence is feasible under $\nu = 1$ as long as \( R_w > \max\{1, R_u\} \), since the {\it Wolbachia}-infected population can only persist when its reproductive fitness exceeds that of the uninfected population. For \( \nu \neq 1 \), coexistence depends on whether \( R_\Delta \geq 1 \iff \Delta \geq 0\), modulated by the signs and ranges of the auxiliary quantities \( S_1 \), \( S_2 \) and \( S_3 \), which capture the relationship between the fitness of uninfected and infected populations under competition.  For a given set of parameters, Figure~\ref{fig:existenceO} shows the regions where one, two, or no coexistence equilibria can be found as $R_w$ varies. Below the red line, $\bar O < k\lambda_i(R_w-1)/c$.  The light cyan,  blue, and pink colors show, respectively, the parameters set for which $f(\bar O)=A_1 \bar O^2 + B_1 \bar O + C_1$ admits two, one, or no real roots. The continuous and dashed black lines represent stable and unstable equilibrium points (real roots of the polynomial), respectively.

\begin{figure}
\centering
\begin{subfigure}{0.48\textwidth}
\includegraphics[width=1\linewidth]{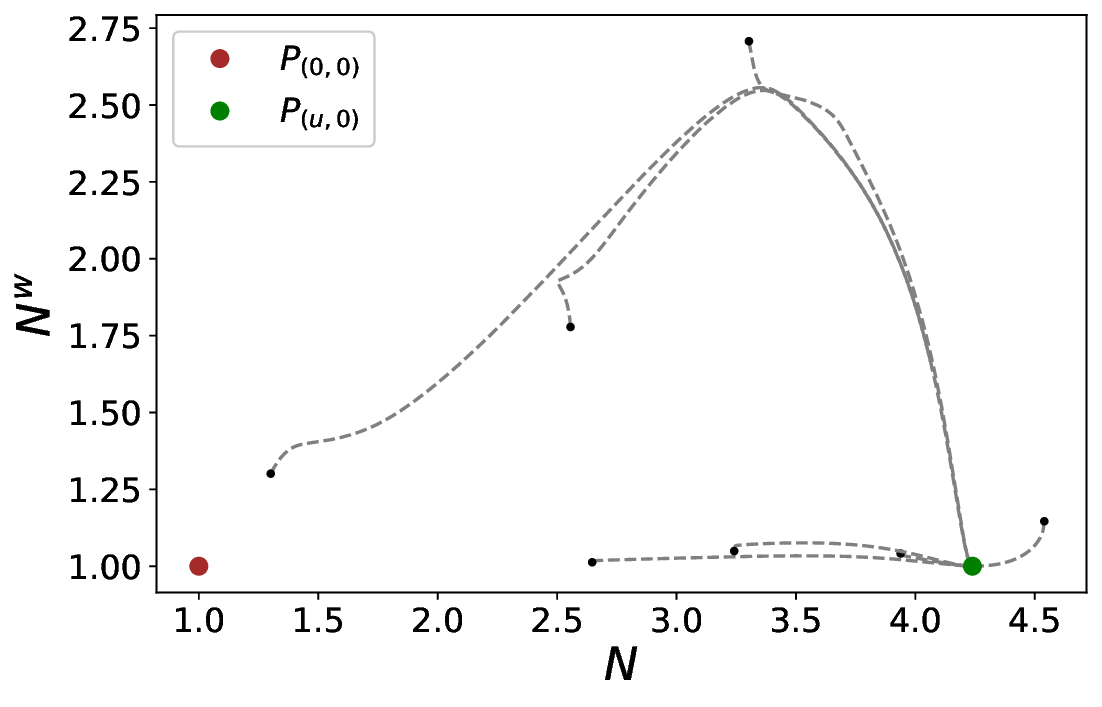}
  \caption{$\eta_o=\eta_o^w=0.12$ and $\zeta=0.8$}
  \label{fig:phase_pu.eps}
\end{subfigure}
\quad
\begin{subfigure}{0.48\textwidth}
\includegraphics[width=1\linewidth]{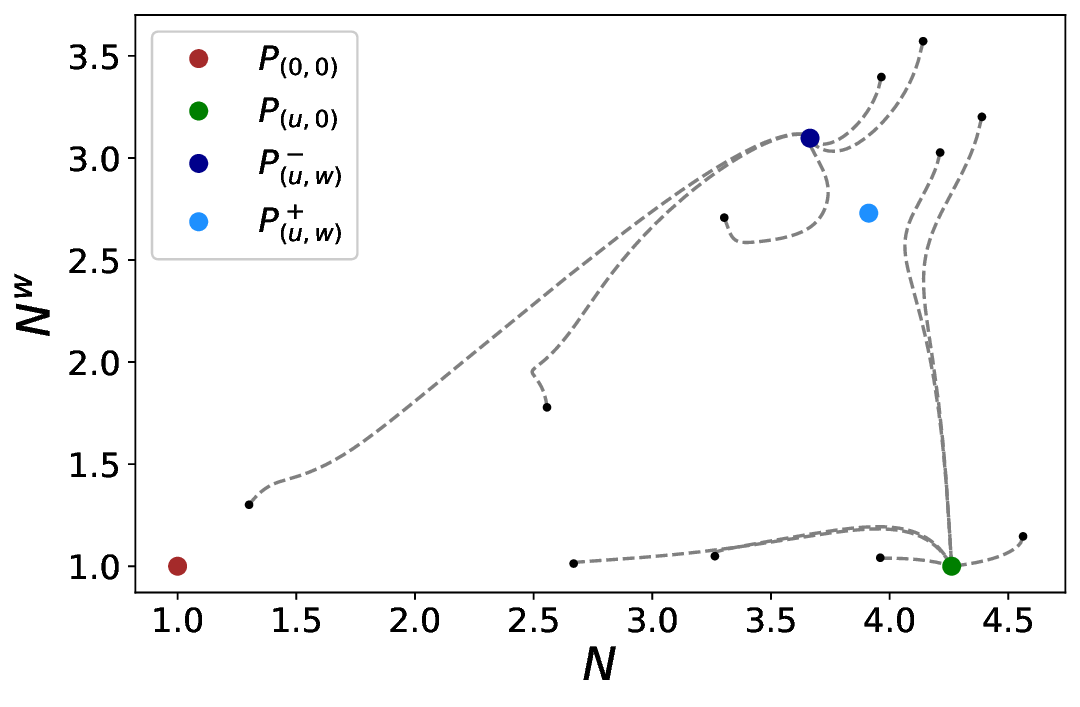}
  \caption{$\eta_o=\eta_o^w=0.09$ and $\zeta=0.8$}
  \label{fig:phase_puw2.eps}
\end{subfigure}
\quad
\begin{subfigure}{0.48\textwidth}
\includegraphics[width=1\linewidth]{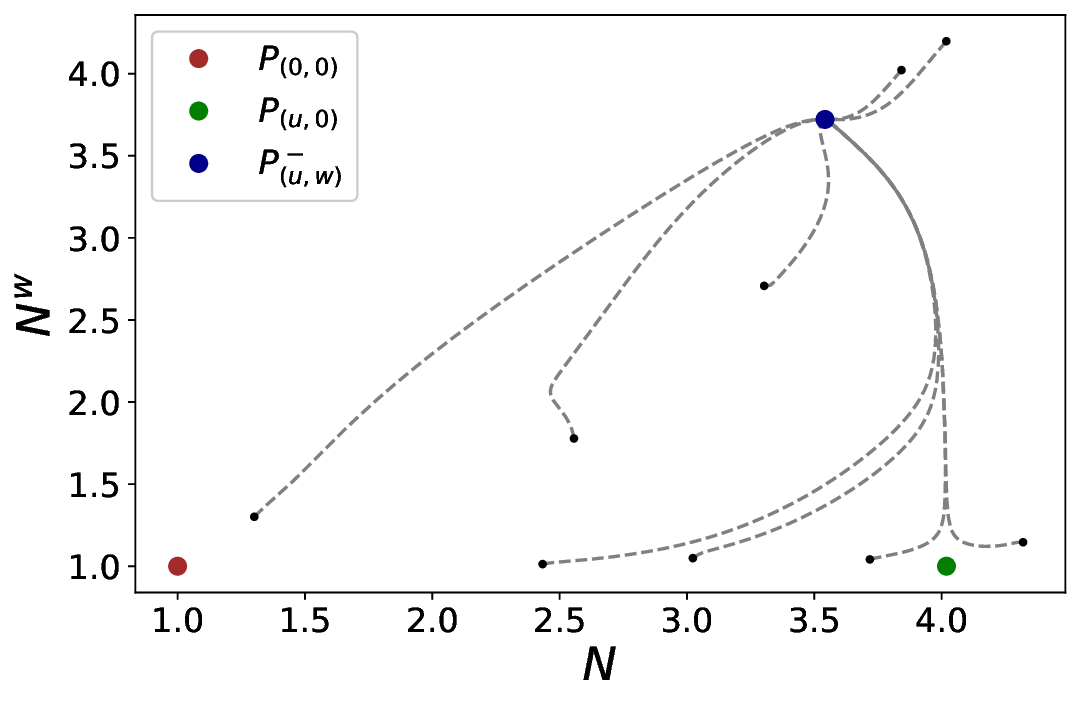}
  \caption{$\eta_o=\eta_o^w=0.01$ and $\zeta=0.8$}
  \label{fig:phase_puw1.eps}
\end{subfigure}
\quad
\begin{subfigure}{0.48\textwidth}
\includegraphics[width=1\linewidth]{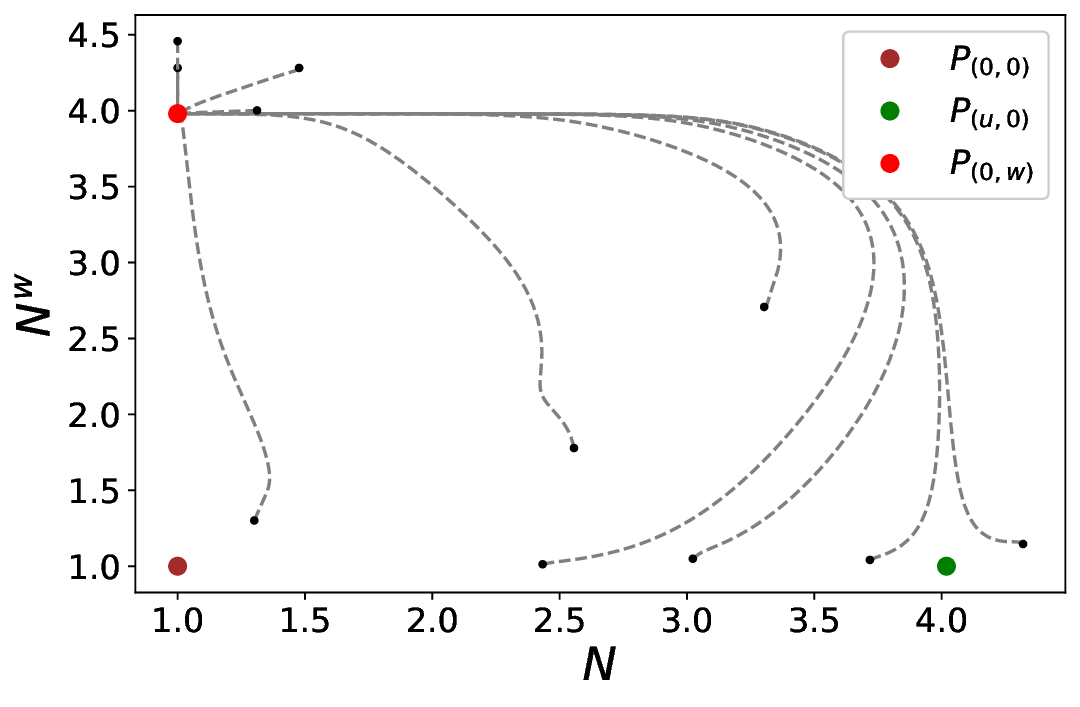}
  \caption{$\eta_o=\eta_o^w=0.01$ and $\zeta=1.0$}
  \label{fig:phase_pw.eps}
\end{subfigure}
\caption{Phase portraits. The vertical axis shows the total uninfected population $\log_{10}(N+10)$ and the horizontal axis shows the total \textit{Wolbachia}-infected population $\log_{10}(N^w+10)$. The dashed lines are different trajectories to the equilibria $P_{(0,0)}, P_{(u,0)}, P_{(0,w)}$ and $P_{(u,w)}$.}
\label{PhasePortraits}
\end{figure}

\begin{figure}
\centering
\begin{subfigure}{.45\textwidth}
  \centering  \includegraphics[width=1\linewidth]{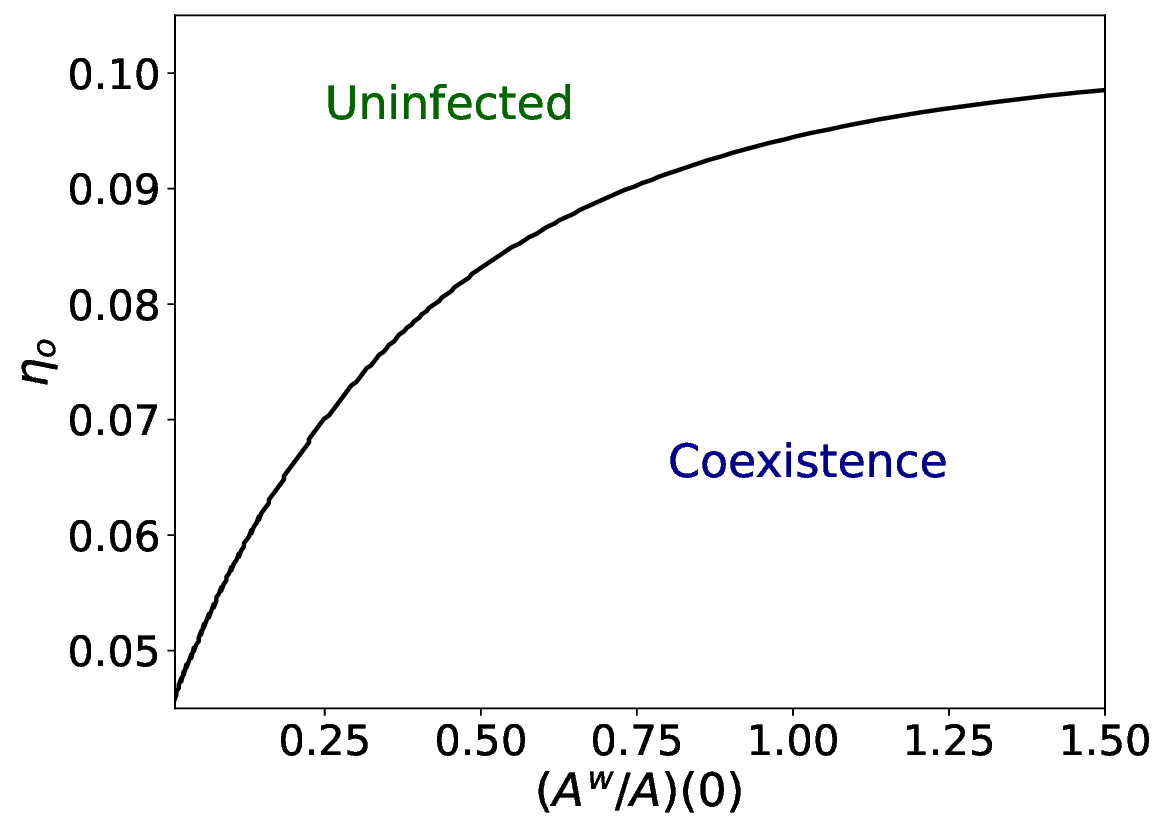}
  \caption{Stability diagram}
  \label{fig:libera}
\end{subfigure}
\begin{subfigure}{.45\textwidth}
 \centering
 \includegraphics[width=1\linewidth]{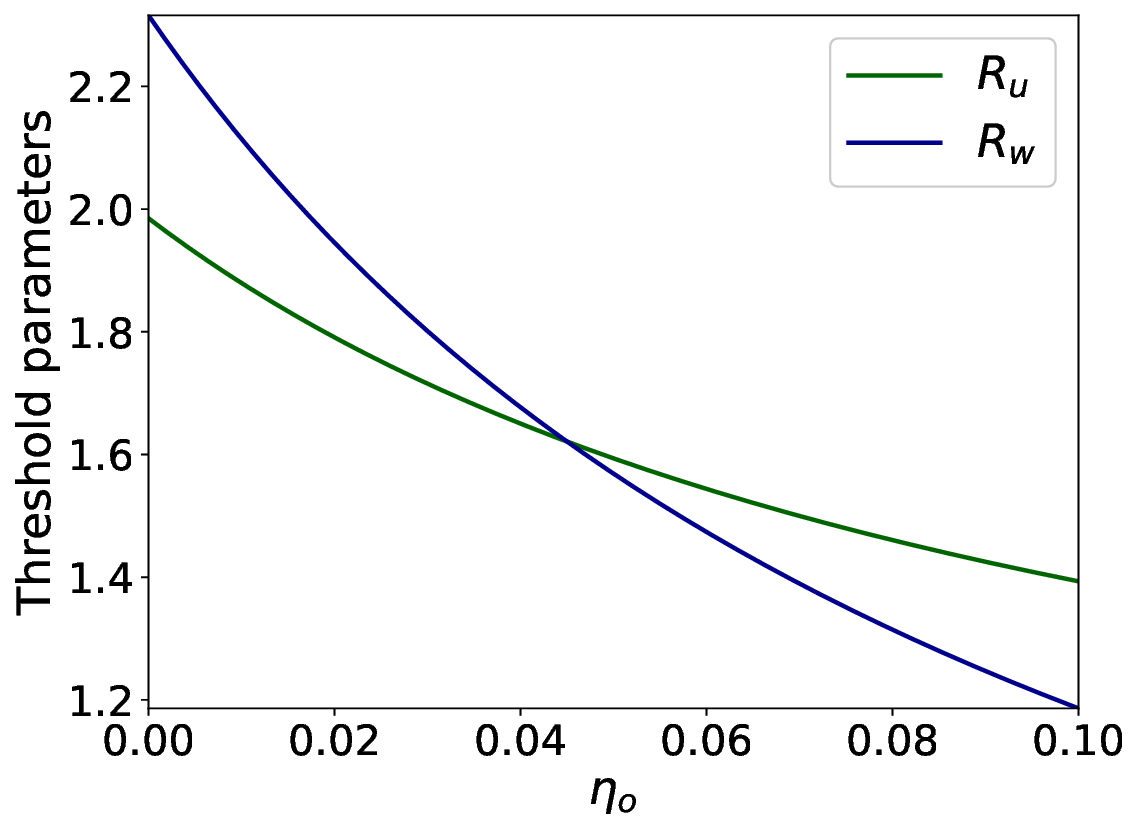}
  \caption{$R_u$ and $R_w$ vs $\eta_o$}
  \label{fig:rwru_eta}
\end{subfigure}
\caption{In (a) the stability diagram showing the asymptotic outcome of mosquito population dynamics as a function of $\eta_o$ and the ratio between the number of \textit{Wolbachia}-infected and uninfected mosquitoes at the moment of the release $(A^w/A)(0)$. In (b) the threshold parameters $R_u$ and $R_w$ are shown as a function of $\eta_o$.}
\label{etaInf}
\end{figure}

Finally, Figure~\ref{PhasePortraits} illustrates typical trajectories in the phase space of the dynamical system with various initial conditions. In all of them, $t=0$, $N(0)>0, N^w(0)>0$ and $(I+I^w)(0)<k$, 
where $N(0)$ and $N^w(0)$ are the total number of uninfected and infected populations, respectively; and $I(0)$ and $I^w(0)$ are the immature uninfected and infected populations, respectively. In each subfigure, the parameters set is varied (relative to the baseline parameters set), and only the existing equilibrium points are shown.  Therefore, the subfigure~\ref{fig:phase_pu.eps} shows the equilibria \( P_{(0,0)} \) and \( P_{(u,0)} \), with the latter stable, indicating failure of the infected strain to establish partially or completely. In subfigure~\ref{fig:phase_puw2.eps}, four equilibria are observed — \( P_{(0,0)} \), \( P_{(u,0)} \), \( P_{(u,w)}^- \), and \( P_{(u,w)}^+ \) — with \( P_{(u,0)} \) and \( P_{(u,w)}^- \) stable and \( P_{(0,0)} \) and \( P_{(u,w)}^+ \) unstable, the latter being a saddle point. The observed bistability indicates that the long-term behavior of the system is sensitive to the initial conditions. Despite the existence of \( P_{(0,0)} \) and \( P_{(u,0)} \), subfigure~\ref{fig:phase_puw1.eps} shows a single stable equilibrium \( P_{(u,w)}^- \), indicating that coexistence is achieved for a wide range of initial conditions. Subfigure~\ref{fig:phase_pw.eps} presents equilibria \( P_{(0,0)} \), \( P_{(u,0)} \), and \( P_{(0,w)} \), showing complete replacement driven by perfect maternal inheritance.  
Lastly, Figure \ref{etaInf} illustrates the influence of egg quiescence on both the minimum release size of \textit{Wolbachia}-infected mosquitoes required to achieve coexistence (subfigure \ref{fig:libera}), and its impact on the threshold parameters $R_u$ and $R_w$ (subfigure \ref{fig:rwru_eta}). The simulations show how interactions between parameters and the amount of {\it Wolbachia}-infected mosquitoes released influence the system’s long-term dynamics, emphasizing the presence of a critical release threshold required for sustained infection persistence.

\subsubsection{Risk of Arbovirus Transmission}

In the simulations, two different scenarios mimic favorable and unfavorable environmental conditions for the mosquito population. The first comprises the baseline scenario with $\phi=0.3$, $\phi^w=0.27$, $\eta_o=\eta_o^w=0.01$, $\mu_q=0.005$, $\mu_a=0.025$,
$\mu_a^w=0.02875$ on days$^{-1}$, and $\zeta=0.95$,
while in the second, the mortality rates of the immature stages increase $\mu_o=\mu_o^ w=0.105$, $\mu_i=\mu_i^w=0.21$, $\eta_o=\eta_o^w=0.1$ on days$^{-1}$, while the development rates decrease $\sigma_o=\sigma_o^w=0.03$ on days$^{-1}$. The other remaining parameters are unchanged. For these two sets of parameters, at all times $t$, given that the minimum number of \textit{Wolbachia}-infected mosquitoes released to establish infection is reached, coexistence is observed. The probability of encountering an uninfected female mosquito in the scenario $s\in\{a,b\}$ is
\[ 
p^s_A = \frac{rA}{rA+r^wA^w},
\] 
and the corresponding odds and odds ratio are given by:
\[
\text{Odds}_A^s = \frac{p^s_A}{1-p^s_A},
\quad \text{and}\quad
\text{OR}_A = \frac{\text{Odds}_A^a}{\text{Odds}_A^b}=\frac{p^a_A(1-p^b_A)}{p^b_A(1-p^a_A)},
\]
respectively. If OR$_A$ is greater than 1, the chance (or odds) of encountering an uninfected female mosquito is greater in scenario $a$ than in scenario $b$. In addition, the increase in the odds from scenario $b$ to $a$ is measured by $(\text{Odds}_A^a-\text{Odds}_A^b)/\text{Odds}_A^b$.
 
\begin{figure}
\centering
\begin{subfigure}{.45\textwidth}
  \centering  
\includegraphics[width=1\linewidth]{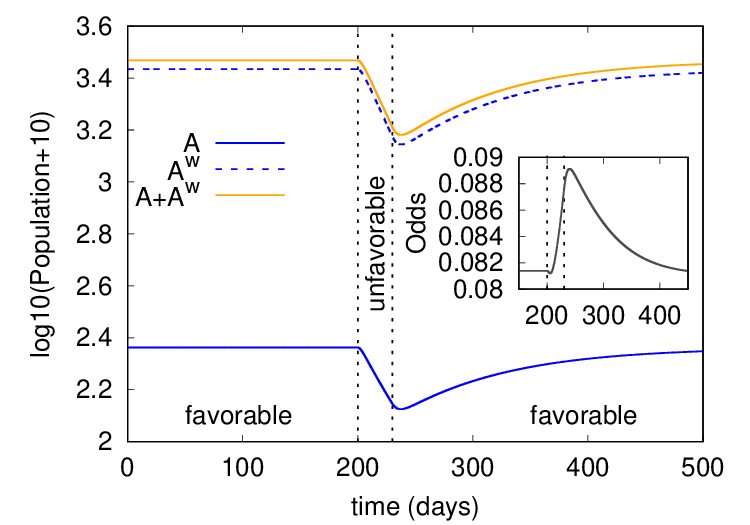}
\caption{Temporal course}
\label{dinamica}
\end{subfigure}
\begin{subfigure}{.45\textwidth}
  \centering  
\includegraphics[width=1\linewidth]{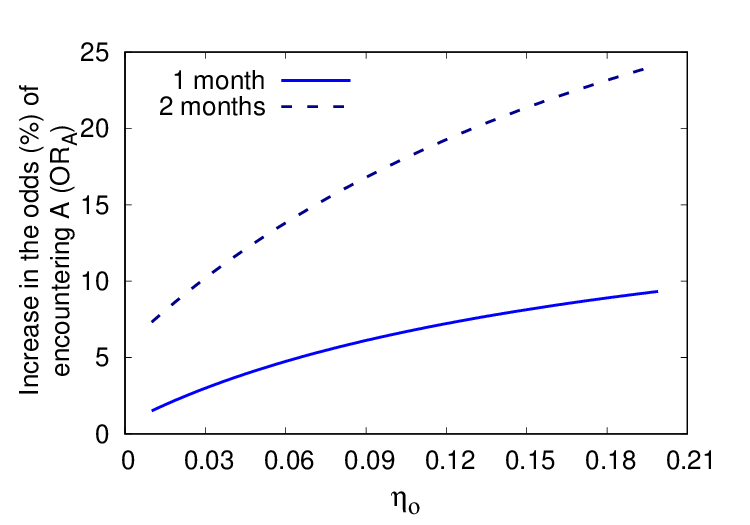}
\caption{Increase in OR$_A$}
\label{risco}
\end{subfigure}
\caption{ On the left, the temporal evolution of the uninfected, infected, and total adult mosquito populations is shown across two distinct periods: a favorable and an unfavorable one. The inset highlights the impact of quiescent eggs on the risk of arbovirus transmission. On the right, the increase in the odds of encountering an uninfected mosquito is presented. The continuous and dashed line corresponds to one and two months of unfavorable period.}
  \label{fig:risco}
\end{figure}

The initial conditions for the unfavorable scenario correspond to the equilibrium values $\bar A$ and $\bar A^w$ in the favorable scenario. During and after the unfavorable period, the probability of encountering an uninfected mosquito changes. 
Figure~\ref{dinamica} shows the temporal evolution of uninfected and \textit{Wolbachia}-infected adult populations, represented by continuous and dashed blue lines, respectively; the total adult population is shown by the orange line. Under unfavorable conditions, the overall mosquito population declines, with a more pronounced reduction in the infected population compared to the uninfected one. Because of it, $p^s_A$ changes in time (as highlighted in the inset panel). Therefore, we evaluate this probability over a $\Delta t$-day interval: the first $\Delta t_1$ corresponds to the unfavorable scenario, and the remaining $\Delta t_2$ to the favorable scenario:
\[
p^s_A=\frac{1}{\Delta t}\int_0^{\Delta t} \frac{rA(t)}{rA(t)+r^wA^w(t)}dt,
\]
and used it to evaluate the odds. The values obtained — ranging from 0.08 to 0.09 for $(\Delta t_1, \Delta t_2) = (30, 60)$, and from 0.08 to 0.105 for $(\Delta t_1, \Delta t_2) = (60, 60)$ — confirm that, under this set of parameters and in both scenarios, the probability of encountering an uninfected mosquito after unfavorable environmental conditions remains low (approximately 7.5\% to 8.3\% or 7.5\% to 9.5\%, respectively). In Figure~\ref{risco}, we vary $\eta_o$ and $\eta^w_o$ simultaneously from 0.01 to 0.2, keeping them equal in each simulation. The continuous and dashed lines correspond to the $(30,60)$ and $(60,60)$ scenarios, respectively. We observe that the odds ratio of encountering an uninfected mosquito increases as $\eta_o$ and $\eta^w_o$ increase. Since mosquitoes that do not carry the bacteria have a higher chance of carrying the dengue virus compared to those infected with \textit{Wolbachia}, and considering the coupling between human and mosquito populations, the risk of dengue virus transmission to the human population may slightly increase during and shortly after the unfavorable period, given that OR$_A>1$. 

\section{Discussion}

The introduction of virus-blocking strains of \textit{Wolbachia} into wild, \textit{Wolbachia}-free \textit{Ae. aegypti} population, followed by the release of \textit{Wolbachia}-infected mosquitoes, has emerged as a cost-efficient and sustainable technique for controlling arbovirus infections. Although infected adult mosquito releases have been carried out in several countries with medium to high success in establishing infection within mosquito populations \citep{tsai2017impact, ryan2019establishment, nazni2019establishment, crawford2020efficient, utarini2021efficacy}, few studies have evaluated the efficacy of these releases in blocking arbovirus transmission. In particular, in Indonesia, \textit{w}Mel prevalence achieves 95.8\% and reduces the incidence of asymptomatic cases of dengue by 77.1\% \citep{anders2018awed}. In Australia, a reduction of 96\% in dengue incidence in the local population was measured after \textit{w}Mel introduction achieves 80\% prevalence \citep{ryan2019establishment}. In Brazil, despite spatial heterogeneity in entomological outcomes with infection prevalence varying between 40 and 80\%, \textit{w}Mel intervention was associated with a reduction of 69\% in dengue incidence. However, this reduction should be interpreted with caution, as dengue incidence was measured during and immediately after the COVID-19 pandemic.

Meanwhile, several works reported the sensitivity of \textit{Wolbachia}-strains to stressed abiotic conditions such as extreme temperature and humidity \citep{ross2017wolbachia, lau2020impacts,
 hien2021environmental}. As the eggs of \textit{Ae. agypti} are laid above the waterline and, especially, in artificial oviposition containers, the absence of water can trigger quiescence in the embryo.  The quiescence period can last six months or more, and quiescent eggs exhibit greater resistance to desiccation (i.e., lower mortality rates) compared to the immature stages (larvae and pupae) of the mosquito life cycle \citep{soares2016eggs, petersen2023dengue}. After contact with water and under favorable abiotic conditions, the egg hatches, and first-instar larvae emerge. Although extreme abiotic conditions generally reduce mosquito fitness, the ability to remain quiescent under such conditions can offer a survival advantage. This dormancy strategy allows eggs to endure environmental stress, increasing the chances of population persistence and facilitating mosquito spread once conditions become favorable again \citep{oliva2018quiescence, JulianoEggs, ByttebierEggs}. However, this is not true for the infected mosquito. In this case, the viability of quiescent \textit{Wolbachia}-infected eggs is reduced, and mosquitoes hatching from it showed partial loss of cytoplasmic incompatibility and female infertility \citep{mcmeniman2010virulent, farnesi2019embryonic,lau2021infertility,ross2020persistent, ross2022fitness}. Although releasing infected mosquitoes is the most common method used in the field, it can be replaced by the release of infected eggs. However, since infected eggs are often stored at low or high temperatures to halt development prior to release, this adds further complexity to the process \citep{allman2023enhancing, WHOwolbachia}.  In \cite{lau2021infertility}, \textit{w}Mel and \textit{w}AlbB infected eggs were stored under cycling temperatures of $11-19^{\circ}$C and $22-30^{\circ}$C for up to 16 weeks. For both bacterial strains, their density tended to decrease with egg storage time, and egg production and viability were more adversely affected by warmer environments. {\it w}Mel-infected males emerging from storage eggs show incomplete CI in both temperature regimes, while {\it w}AlbB-infected females suffered infertility when stored as eggs at $22-30^{\circ}$C.

Considering the quiescence mechanisms by which \textit{Ae. aegypti} eggs can survive extreme abiotic conditions, we developed an ordinary differential equation model to investigate their impact on the prevalence of \textit{Wolbachia} infection in mosquito populations. In the model, quiescent infected eggs are assumed not to contribute to the emergence of adult infected mosquitoes. Although we treat quiescence as a constant rate process, it is likely to depend on environmental conditions. Analytical and numerical results, obtained from the proposed model, show that, when only one population — either uninfected or \textit{Wolbachia}-infected — is present in the environment, the population dynamics is determined by the parameters $R_u$ and $R_w$, which represent the reproductive fitness of uninfected and infected populations in isolation, as well as by the initial number of infected mosquitoes released. On the other hand, when uninfected and infected mosquitoes coexist, their competitive dynamics are characterized by the parameters \( R_\Delta \), \( S_1 \), \( S_2 \), and \( S_3 \), defined in terms of \( R_u \), \( R_w \), and \( R_{uw} \), where the latter measures the number of uninfected females generated by an \textit{Wolbachia}-infected female due to imperfect inheritance. In all cases, \textit{Wolbachia} infection can establish itself in the mosquito population only if $R_w > 1$ (see Table~\ref{Table_rurw}). A practical implication of this result is that the long-term success of releasing infected mosquitoes depends on the \textit{Wolbachia} bacteria strain. Assuming that the mosquito population persists, the outcome — whether the infected population fully replaces the uninfected one (\( \zeta = 1 \)) or population coexistence (\( \zeta \neq 1 \)) — is determined by the initial conditions and the parameter values. In general, higher rates of quiescence affect the establishment and prevalence of infection in mosquito populations, highlighting the need for a higher number of infected mosquitoes to be released to ensure successful \textit{Wolbachia} establishment (Figs.~\ref{BifurcationDiagrams}, \ref{RuRwcase}, \ref{fig:existenceO}, \ref{PhasePortraits}, and \ref{etaInf}). 
 
Regarding the effect of quiescent eggs on the release of \textit{ Wolbachia}-infected mosquitoes, the sensitivity analysis indicates a relationship between quiescence rates, mortality rates, and development rates during the egg stage of the mosquito life cycle. Although temperature changes negatively impact mosquito fitness regardless of infection status (see Fig.~\ref{dinamica}), quiescence has a rescue effect. This is because quiescent eggs are more resistant to extreme temperatures and humidity (which includes lack of water). Therefore, the mosquito population must optimize its fitness by adjusting the quiescence rate (see the relationship between $R_u$, $R_{uw}$, and the parameters related to the egg stage). In contrast, $R_w$ always decreases with increasing $\eta_o^w$, indicating that quiescence impacts the prevalence and persistence of the \textit{Wolbachia}-infected population due to higher mortality and infertility (see Figs.~\ref{fig:P_u0_to_Puw} and \ref{etaInf}). Because of this, the chance that an uninfected mosquito bites a human increases during and shortly after unfavorable periods, and this increase is positively correlated with the quiescence rate (see Fig. \ref{fig:risco}). This is a consequence of the sporadic decline in \textit{Wolbachia}-infection prevalence in the mosquito population. As uninfected mosquitoes are more susceptible to the dengue virus, this may increase dengue transmission within the human-mosquito population. However, it is important to note that the main effect of quiescence is related to the difficulty in achieving \textit{Wolbachia}-infection establishment within the mosquito population, as the minimum number of infected mosquitoes required for release increases with the quiescence rate (Fig. \ref{fig:libera}). Nonetheless, the potential loss of the infection from the population due to quiescence should not be disregarded (Fig. \ref{fig:P_u0_to_Puw}).

Therefore, although the model developed here emphasizes that \textit{Wolbachia} infection prevalence in the mosquito population decreases during unfavorable environmental conditions that change mosquito entomological parameters, coupling the dynamics of uninfected and \textit{Wolbachia}-infected mosquito populations with dengue virus transmission between humans and mosquitoes is crucial to assess the risk of an increase in dengue cases, as well as the implications for the epidemiology and control of arboviral infections through \textit{Wolbachia}-infected mosquito release.

\section{Conclusion}
Although the size of the mosquito population decreases under unfavorable environmental conditions, the risk of arbovirus transmission may slightly increase, as the likelihood of humans being bitten by uninfected (and potentially virus-carrying) mosquitoes increases. This is because the prevalence of \textit{Wolbachia} infection decreases in areas where mosquito breeding habitats are intermittent, requiring extended periods of egg quiescence, which affects infected mosquitoes more than uninfected ones. More importantly, quiescence can compromise the establishment of infection within the mosquito population. Therefore, quiescent eggs can jeopardize the efficiency of \textit{Wolbachia}-infected mosquito release strategies aimed at reducing the uninfected population and interrupting arbovirus transmission.  As infection traits and mosquito fitness depend on the bacterial strain used during release, it is important to think of a portfolio of \textit{Wolbachia} strains for application as public health interventions.

\section*{Acknowledgments}
The authors would like to express their gratitude in memory of Sergio Muniz Oliva Filho, whose insights helped inspire this work. Artificial Intelligence (OpenAI's ChatGPT) was used to review the English language, correct orthography, and improve reading flow.

\section*{Funding}
LESL acknowledges support from CAPES – Finance Code 001 – and the Institutional Program for Doctoral Research Abroad (PDSE) for the scholarship. CPF thanks the financial support from grant \#304007/2023-4 from CNPq. This work was supported by grants \#2019/22157-5, \#2020/10964-0, \#21/09004-5, São Paulo Research Foundation (FAPESP), and CAPES \#88881.878875/2023-01. 

\input{main.bbl}

\newpage{}

\begin{appendices}

\section{Proofs}  \label{proof1}
Here, the proofs of Theorems \ref{theorem1} and \ref{theorem2} are given.\\

\noindent{\bf Theorem 1.} If $X(0)>0 \mbox{ and } (I+I^w)(0)<k$,  then, for all $t>0$,  $X(t)$ exits, it is unique, and satisfies 
\begin{eqnarray}
X(t)>0 \mbox{ and } (I+I^w)(t)<k. \label{positivity}
\end{eqnarray}

\begin{proof}  From the existence and uniqueness theorem, since $X(0)>0$ and from the regularity of system \eqref{systemO}, we have that $X(t)$ exists and is unique as long as $X(t)$ is positive. So the only thing left to prove is \eqref{positivity}.

Using the method of variation of constants, we obtained that
\begin{eqnarray}
A(t)&=&A(0)e^{-\mu_at}+\sigma_ie^{-\mu_at}\int_0^te^{\mu_as}I(s)ds,\nonumber\\  
A^w(t)&=&A^w(0)e^{-\mu_a^wt}+\sigma_i^we^{-\mu_a^wt}\int_0^te^{\mu_a^ws}I^w(s)ds,\label{posit}\\
Q(t)&=&Q(0)e^{-(\sigma_q+\mu_q)t}+\eta_oe^{-(\sigma_q+\mu_q)t}\int_0^te^{(\sigma_q+\mu_q)s}O(s)ds, \nonumber\\
O(t)&=&O(0)e^{-(\sigma_0+\eta_0+\mu_0)t}+e^{-(\sigma_0+\eta_0+\mu_0)t}\int_0^te^{(\sigma_0+\eta_0+\mu_0)s}\left[\dots\right],\nonumber\\
O^w(t)&=&O^w(0)e^{-(\sigma^w_0+\eta^w_0+\mu^w_0)t}+e^{-(\sigma^w_0+\eta^w_0+\mu^w_0)t}\int_0^te^{(\sigma^w_0+\eta^w_0+\mu^w_0)s}\phi^w \zeta r^wA^w(s)ds. \nonumber    
\end{eqnarray}

The proof follows by contradiction, assuming that \eqref{positivity} is false. From the hypothesis ($X(0)>0 \mbox{ and } (I+I^w)(0)<k$) and the continuity of solutions, we have that there exists $t_1>0$ such that $X(t)>0$, $(I+I^w)(t)<k,~ \forall~ 0\leq t < t_1,$ and \eqref{positivity} is false for $t=t_1.$

One can easily see that $\{A(t_1),~ Q(t_1),~ A^w(t_1),~O(t_1),~O^w(t_1)\}>0$. 

Therefore, if $(I+I^w)(t_1)=k$ then either $I(t_1)>0$ or $I^w(t_1)>0$ and in both cases we get $\left.\frac{d}{dt}(I+I^w)\right|_{t_1}<0$, which is a contradiction. Thus, $(I+I^w)(t_1)<k$.

Finally, assuming  that $I(t_1)=0$ (respectively, $I^w(t_1)=0$), since $O(t_1)>0$ and $Q(t_1)>0$ (respectively, $O^w(t_1)>0$), we get that $\left.\frac{dI}{dt}\right|_{t_1}>0$ (respectively, $\left.\frac{dI^w}{dt}\right|_{t_1}>0$), another contradiction since $I(t)>0 \mbox{ (respectively, } I^w(t)>0),~ 0\leq t < t_1.$

Which proves that $X(t)>0 \mbox{ and } (I+I^w)(t)<k  \mbox{ for all } t > 0.$
\end{proof}

One can relax the hypotheses of the previous theorem and prove the same result assuming that $O(0),~ A(0),~ Q(0),~ O^w(0)$ and $ A^w(0)$ are non-negative, $I(0)$ and $I^w(0)$ are strictly positive, and $(I+I^w)(0)<k$. The proof follows immediately noticing that, from the continuity of solutions and \eqref{positivity}, there exists $\delta>0$  such that $X(\delta)>0 \mbox{ and } (I+I^w)(\delta)<k.$\\

\noindent{\bf Theorem 2.} Assume that $X(0)>0 \mbox{ and } (I+I^w)(0)<k$.  Then, there are constants $\tau>0$ and $\tau_w>0$ such that, for all $t>0$, we have  $N(t) \leq \tau$ and $N^w(t) \leq \tau_w.$ \label{bondedness}
\begin{proof}
From system \eqref{systemO}, we can see that $N(t)=O(t)+I(t)+A(t)+Q(t)$ and $N^w(t)=O^w(t)+I^w(t)+A^w(t)$  satisfy
\begin{eqnarray}
\frac{dN(t)}{dt}&=& \phi rA(t) \left[\frac{(1-r)A+ \nu(1-r^w)A^w}{(1-r)A+(1-r^w)A^w}\right] +  \phi^w r^wA^w(1-\zeta) + 
\nonumber \\
&&- \mu_oO  - \mu_iI -\mu_a A-\mu_qQ-(\sigma_o O + \sigma_q Q) \left(\frac{I(t)+I^w(t)}{k}\right),\nonumber \\
\frac{dN^w(t)}{dt}&=& \phi^w \zeta r^wA^w(t) - (\eta_o^w+ \mu_o^w)O^w(t) - \mu_i^wI^w(t) +  \nonumber \\
&&- \mu_a^wA^w(t) -\sigma_o^w O^w(t) \left(\frac{I(t)+I^w(t)}{k}\right).\nonumber 
 \end{eqnarray}
As, for all $t \geq 0$, we have that
\begin{eqnarray}
    A(t)&=&A(0)e^{-\mu_at}+\sigma_ie^{-\mu_at}\int_0^te^{\mu_as}I(s)ds \quad \mbox{with} \quad I<k,\quad \mbox{and}\nonumber\\ 
    A^w(t)&=&A^w(0)e^{-\mu_a^wt}+\sigma_i^we^{-\mu_a^wt}\int_0^te^{\mu_a^ws}I^w(s)ds \quad \mbox{with} \quad I^w<k, \nonumber    
\end{eqnarray}
then $A(t) \leq A(0)+\frac{\sigma_i}{\mu_a}k$, $A^w (t)\leq A^w(0)+\frac{\sigma_i^w}{\mu_a^w}k$. Also note that $\frac{(1-r)A}{(1-r)A+(1-r^w)A^w} \le 1$ and $\frac{\nu (1-r^w) A^w(t)}{(1-r)A+(1-r^w)A^w} \le 1$. Therefore,
\begin{eqnarray}
&&\frac{dN(t)}{dt}\leq 2\phi r\left(A(0)+\frac{\sigma_i}{\mu_a}k\right)+\phi^w r^w(1-\zeta)\left(A^w(0)+\frac{\sigma_i^w}{\mu_i^w}k\right)
-\gamma_1 N(t)\nonumber \\
&&\frac{dN^w(t)}{dt}\leq \phi^w r^w\zeta \left(A^w(0)+\frac{\sigma_i^w}{\mu_i^w}k\right) -\gamma_2 N^w(t) \label{NNw1}
 \end{eqnarray}
 where $\gamma_1=\min\{\mu_o, \mu_i,\mu_a, \mu_q\}$ e $\gamma_2=\min\{\eta_o^w+ \mu_o^w, \mu_i^w, \mu_a^w\}$.

Observe that the first-order linear system given by
 \begin{eqnarray}
     \frac{dZ(t)}{dt}&=&\rho_1+\rho_2-\gamma_1Z (t),\label{NNw2}\\
     \frac{dZ^w(t)}{dt}&=&\rho_3-\gamma_2Z^w(t), \nonumber
 \end{eqnarray}
with $\rho_1,\rho_2, \rho_3,\gamma_1$ and $\gamma_2$ strictly positive constants, can be solved analytically, i.e.,
 \[
     Z(t)=Z(0)e^{-\gamma_1t}+\frac{\rho_1+\rho_2}{\gamma_1}(1-e^{-\gamma_1t}), \quad
     Z^w(t)=Z^w(0)e^{-\gamma_2t}+\frac{\rho_3}{\gamma_2}(1-e^{-\gamma_2t}). 
 \]
 Therefore, from \eqref{NNw1} and \eqref{NNw2} we can conclude that
 \begin{eqnarray}
    && 0\le N(t)\le N(0)e^{-\gamma_1t}+\frac{\rho_1+\rho_2}{\gamma_1}(1-e^{-\gamma_1t}), \quad \mbox{and}\nonumber\\
    && 0\le N^w(t)\le N^w(0)e^{-\gamma_2t}+\frac{\rho_3}{\gamma_2}(1-e^{-\gamma_2t}), \nonumber
 \end{eqnarray}
 where\begin{eqnarray}
\rho_1&=&2\phi r\left(A(0)+\frac{\sigma_i}{\mu_a}k\right), \nonumber\\
\rho_2&=& \phi^w r^w(1-\zeta)\left(A^w(0)+\frac{\sigma_i^w}{\mu_i^w}k\right),\nonumber\\
\rho_3&=&\rho_2\frac{\zeta}{1-\zeta}.\nonumber
\end{eqnarray}
 This implies that
 \[
     N(t) \le N(0)+\frac{\rho_1+\rho_2}{\gamma_1} \quad \mbox{and} \quad 
     N^w(t) \leq N^w(0)+\frac{\rho_3}{\gamma_2}. 
 \]
In other words, $N(t)$ and $N^w(t)$ are bounded
and 
\[
\tau = N(0)+\frac{\rho_1+\rho_2}{\gamma_1}\quad \mbox{and}\quad \tau_w=N^w(0)+\frac{\rho_3}{\gamma_2}.
\]

\end{proof}

\section{Equilibrium Points} \label{existence_equilibrium}

The equilibrium points - $(\bar O, \bar I, \bar A, \bar Q, \bar O^w, \bar I^w,\bar A^w)$ - correspond to the time-independent solutions of system \eqref{systemO}, and they  are obtained by solving the nonlinear system given by 
\begin{eqnarray}
&&0=\phi r \bar A \left(\frac{\underline r\bar A + \nu \underline r^w\bar A^w}{\underline r \bar A+ \underline r^w\bar A^w}\right)+ \phi^w r^w \underline \zeta \bar A^w  -\lambda_o\bar O \label{eq10}\\
&&0=(\sigma_o \bar O + \sigma_q \bar Q) \left(1- \frac{\bar I+\bar I^w}{k}\right) - \lambda_i\bar I \label{eq20} \\
&&0= \sigma_i \bar I-\mu_a \bar A\label{eq30}\\
&&0= \eta_o \bar O- \lambda_q\bar Q \label{eq40}\\
&&0= \phi^w r^w  \zeta \bar A^w - \lambda_o^w \bar O^w\label{eq50}\\
&&0= \sigma_o^w \bar O^w \left(1- \frac{\bar I+\bar I^w}{k}\right)- \lambda_i^w\bar I^w\label{eq60}\\
&&0= \sigma_i^w \bar I^w- \mu_a^w \bar A^w.\label{eq70}
 \end{eqnarray}
Firstly, observe that  
\[
\frac{\underline r\bar A}{\underline r \bar A+ \underline r^w\bar A^w} \le 1 \quad \mbox{and} \quad \frac{\nu\underline r^w\bar A^w}{\underline r \bar A+ \underline r^w\bar A^w} \le 1.
\]
Therefore, when $\bar A\longrightarrow 0$ and $\bar A^w\longrightarrow 0$, we have
\[
\bar A \left(\frac{\underline r\bar A + \nu \underline r^w\bar A^w}{\underline r \bar A+ \underline r^w\bar A^w}\right) \longrightarrow 0
\]
which gives the trivial equilibrium 
\[
P_{(0,0)}=(0,0,0,0,0,0,0),
\]
where both mosquito populations go to extinction. To obtain the other ones, we suppose that either $\bar A \ne 0$ or $\bar A^w \ne 0$.
Then, after some algebraic manipulation, we obtain from Eqs. \eqref{eq40} and \eqref{eq60}
 \begin{equation}
 \bar Q=\frac{\eta_o }{\lambda_q} \bar O, \quad 
\bar I^w=\frac{\sigma_o^w \bar O^w(k-\bar I)}{\lambda_i^w k+\sigma_o^w \bar O^w}\quad \mbox{with}\quad    \bar I<k.
\label{Iweq}
 \end{equation}
In the same way, from Eqs. \eqref{eq30} and  \eqref{eq70} we have
 \begin{equation}
 \bar A=b\bar I, \quad
\bar A^w=b^w\bar I^w \quad \mbox{with} \quad b:=\frac{\sigma_i}{\mu_a} \quad \mbox{and} \quad  b^w:=\frac{\sigma^w_i}{\mu^w_a}.\label{Aweq}
\end{equation}
 Substituting  $\bar Q$ and $\bar I^w$ into Eq. \eqref{eq20}
 \begin{equation}
 \bar I= \frac{c \bar O \lambda_i^wk}{  \lambda_i (\lambda_i^w k+\sigma_o^w \bar O^w) + \lambda_i^wc\bar O} \quad \mbox{with} \quad c:=\sigma_o+\frac{\sigma_q\eta_o}{\lambda_q}.
 \label{Ieq}
\end{equation}
From Eq. \eqref{eq50} 
\begin{equation}
\left(\frac{d^w \sigma_o^ w(k-\bar I)}{\lambda_i^wk+\sigma_o^w \bar O^w}-\lambda_o^w\right) \bar O^w=0 \quad \mbox{with} \quad  \quad d^w:=\phi^w \zeta r^wb^w.\nonumber
\end{equation}
From the last expression, we can see that $\bar O^w=0$ or
$
\displaystyle{\left(\frac{d^w\sigma_o^w(k-\bar I)}{\lambda_i^w k +\sigma_o^w \bar O^w}-\lambda_o^w\right)=0}.
$
Therefore, we have two cases:
\begin{enumerate}
    \item[(i)] case $\bar O^w=0$. Given that all parameters of the model are positive, if one component of the infected mosquito population is zero, the others are too. Thus,
    $\bar I^w=\bar A^w=0$.
\end{enumerate}
Setting $\bar A^w=0$ into Eq. \eqref{eq10} 
\[
\phi r \bar A - \bar O\lambda_o=0.
\label{sitII}
\]
Substituting the expressions of $\bar A$ and $\bar I$ obtained before (Eqs. \eqref{Aweq} and \eqref{Ieq} with $\bar O^w=0$) in the equation above, we have
\[
\bar O\left(\frac{\phi r b ck}{\lambda_i k  +c \bar O } - \lambda_o\right)=0.
\]
Therefore,
\[
\bar O= \frac{\phi r b c  k- \lambda_o\lambda_ ik}{c\lambda_o},
\label{Oeq}
\]
since we are looking for other solutions than $P_{(0,0)}$.
The strict positive of $\bar O$ is ensured by
\[
\phi r b c  - \lambda_o\lambda_i>0 \Longleftrightarrow R_u>1; \quad R_u:=\frac{\phi r b c}{\lambda_o\lambda_i}=\frac{\phi r \sigma_i c}{\lambda_o\lambda_i\mu_a}.
\]

Therefore, the equilibrium is given by
\[
P_{(u,0)}=(\bar O, \bar I, \bar A, \bar Q,0,0,0),
\]
and corresponds to the extinction of the infected population and the persistence of the uninfected population. Observe that $P_{(0,0)}$ always exists, but the existence of $P_{(u,0)}$ is guaranteed by $R_u>1$.

In summary, the components of $P_{(u,0)}$ as a function of $R_u$ are given by
\[
\begin{aligned}[t]
\bar O&= \frac{k\lambda_i (R_u-1)}{c}, ~
\bar I=\frac{k(R_u-1)}{R_u}, ~
\bar A=\frac{bk(R_u-1)}{R_u}, ~\mbox{and}~
\bar Q= \frac{\eta_ok\lambda_i(R_u-1)}{\lambda_qc}.
\end{aligned}
\]

\begin{enumerate}
\item[(ii)] case 
\[
\left(\frac{d^w\sigma_o^w(k-\bar I)}{\lambda_i^w k +\sigma_o^w \bar O^w}-\lambda_o^w\right)=0
\label{caso2}
\]
\end{enumerate}

 By substituting $\bar I$ given by Eq. \eqref{Ieq} into the the equation above we obtain
\begin{eqnarray}
  && d^w\sigma_o^w\left(k-\frac{c \bar O \lambda_i^wk}{  \lambda_i (\lambda_i^w k+\sigma_o^w \bar O^w) + \lambda_i^wc\bar O}\right)= \lambda_o^w(\lambda_i^w k +\sigma_o^w \bar O^w) \nonumber \\ 
 &&\bar O^w =\frac{d^w\sigma_o^w k\lambda_i - \lambda_o^w (\lambda_i \lambda_i^w k + \lambda_i^wc\bar O)}{\lambda_o^w\lambda_i \sigma_o^w }. 
\end{eqnarray}

Now, we can rewrite all the variables (Eqs. \eqref{Iweq}, \eqref{Aweq}, and \eqref{Ieq}) as functions of $\bar{O}$, as follows:
\begin{eqnarray}
 &&  \bar Q=\frac{\eta_o }{\lambda_q} \bar O, \quad \bar I= \frac{\lambda_o^w c  \lambda_i^w\bar O}{d^w\sigma_o^w \lambda_i}, \quad
\bar I^w =\frac{d^w\sigma_o^w k\lambda_i - \lambda_o^w \lambda_i^w(k \lambda_i + c\bar O)}{d^w\sigma_o^w \lambda_i}, \nonumber \\
&& \bar A = b\frac{\lambda_o^w c  \lambda_i^w\bar O}{d^w\sigma_o^w \lambda_i}, \quad \mbox{and}  \quad \bar A^w = b^w \frac{d^w\sigma_o^w k\lambda_i - \lambda_o^w \lambda_i^w(\lambda_i k + c\bar O)}{d^w\sigma_o^w \lambda_i}. \label{COEX}
\end{eqnarray}
Substituting $\bar O=0$ into Eq. \eqref{COEX} and $\underline \zeta = 0$ (i.e. $\zeta =1)$ into  Eq.~\eqref{eq10},  we obtain  
\[
P_{(0,w)}=(0,0,0,0,\bar O^w, \bar I^w, \bar A^w),
\]
with 
\[
\bar O^w= \frac{k\lambda_i^w(R_w-1)}{\sigma_o^w}, \quad  \bar I^w =\frac{k(R_w-1)}{R_w}, \quad  \bar A^w = \frac{b^wk(R_w-1)}{R_w},
\]
and
\[
R_w:=\frac{d^w\sigma_o^w}{\lambda_o^w\lambda_i^w}=\frac{\phi^w r^w \zeta \sigma_i^w }{\lambda_o^w\lambda_i^w\mu_a^w}.
\]
This equilibrium exists if and only if $R_w > 1$ and $\zeta = 1$, and corresponds to the persistence of the infected mosquito population and the extinction of the uninfected mosquito population.

The last equilibrium 
\[
P_{(u,w)}=(\bar O,\bar I,\bar A,\bar Q,\bar O^w, \bar I^w, \bar A^w)
\]
corresponds to the persistence of both mosquito populations with $\zeta\neq 1$. To obtain it, we substitute the $\bar{I}$, $\bar{I}^w$, $\bar{A}$, $\bar{A}^w$, and $\bar{O}^w$ given by Eq.~\eqref{COEX} into Eq.~\eqref{eq10} and obtain:

\begin{equation}
    f(\bar O)=A_1\bar O^2+B_1\bar O+C_1=0 \label{eqO2}
\end{equation}where
\begin{eqnarray}
    A_1 &=& -c [\underline r b (R_w+R_{uw}-R_u)-\underline r^w b^w (R_w+R_{uw}-\nu R_u)],\nonumber\\
    B_1 &=& k \lambda_i(R_w -1) [\underline r b R_{uw} -\underline r^w b^w(R_w +2R_{uw}-\nu R_u)], \nonumber \\
    C_1 &=&\frac{(k\lambda_i)^2(R_w- 1)^2}{c}\underline r^wb^wR_{uw}>0; \quad R_{uw}:=\frac{\phi^w r^wb^w\underline \zeta c}{\lambda_o\lambda_i}.\nonumber
\end{eqnarray}
Besides, from Eq. \eqref{COEX}, we can see that 
\begin{equation}
    0<\bar O < \frac{k\lambda_i(R_w-1)}{c} \label{restO}
\end{equation} 
which implies that $R_w>1$. In particular, the maximum value of $\bar O$ is obtained when only the uninfected population persists, and it is given by $\bar O = k \lambda_i (R_u - 1)/c$. Rearranging Eq.~\eqref{restO}, we obtain $1 < R_u < R_w$ (i.e. $R_w>\max\{1,R_u\}$). For the general case we have to analysis the discriminant of Eq. \eqref{eqO2} which is given by 
\[\Delta = [k\lambda_i(R_w-1)]^2\left[(\underline r b R_{uw})^2+2\underline r b \underline r^w b^wR_{uw}(R_w +(\nu -2)R_u)+(\underline{r}^w b^w)^2(R_w-\nu R_u)^2\right].\]

This expression can be rewritten in terms of \(\nu\) as:
\begin{enumerate}
    \item [1.] If \( \nu = 1 \), then $\Delta = k^2 \lambda_i^2 (R_w - 1)^2 \left[ \underline{r}b R_{uw} + \underline{r}^w b^w (R_w - R_u) \right]^2\ge 0$. Moreover, if $\zeta \neq 1$ and $R_w>\max\{1,R_u\}$, there exists a unique positive coexistence equilibrium with:
        \[
        \bar{O} = \frac{k\lambda_i(R_w - 1)}{c}\frac{R_{uw}}{R_w + R_{uw} - R_u}.
        \]
    \item [2.] Otherwise, if \( \nu \neq 1 \), then $\Delta = k^2 \lambda_i^2 (R_w - 1)^2 \left[ 4 \underline{r}b \underline{r}^w b^w R_{uw} R_u (1 - \nu)(R_{\Delta} - 1) \right],$
    where
    \[
    R_{\Delta} = \frac{ \left[ \underline{r} b R_{uw} + \underline{r}^w b^w (R_w - \nu R_u) \right]^2 }{ 4 \underline{r} b \underline{r}^w b^w R_{uw} R_u (1 - \nu)}\quad \mbox{and}\quad R_{\Delta} \geq 1 \iff \Delta \geq 0.
    \]
\end{enumerate}
In summary, the coexistence equilibrium exists  if  $\zeta\neq 1$, $R_w>1$, $\Delta \ge 0$ and \(\bar O\) falls within the admissible range given by Eq. \eqref{restO}. From this point onward, we shall assume that these hypotheses hold.

Applying Descartes's rule of signs to Eq.~\eqref{eqO2}, we obtain:

\begin{remark}
Assuming that $\nu \neq 1$. Then, \label{remark1}
\begin{enumerate}
    \item[(a)] If \(A_1 > 0\) and \(B_1 < 0\), the equation has either one (if \(\Delta = 0\)) or two (if \(\Delta > 0\)) positive real roots. If \(B_1 > 0\), there is no positive real root.
    \item[(b)] If \(A_1 < 0\), then \(\Delta > 0\) and the equation admits a unique positive real root.
\end{enumerate}
\end{remark}

Let us define:
\begin{eqnarray}
    S_1 &:=& \underline{r} b (R_w + R_{uw} - R_u) - \underline{r}^w b^w (R_w + R_{uw} - \nu R_u), \nonumber \\
    S_2 &:=&  \underline{r} b R_{uw} - \underline{r}^w b^w (R_w + 2R_{uw} - \nu R_u), \nonumber \\
    S_3 &:=& 2\sqrt{\underline{r} b \, \underline{r}^w b^w \, R_{uw} R_u (1 - \nu)(R_{\Delta} - 1)}, \quad \mbox{with}\quad R_{\Delta} \geq 1. \nonumber
\end{eqnarray}

These allow us to simplify the expressions and determine the number and nature of positive equilibria. Assuming that $\nu \neq 1$. Then, several scenarios can be drawn:

        \begin{enumerate}
            \item[2.1.] Suppose that $\underline rb = \underline r^w b^w$ and $S_2 < 0 \iff  R_w + R_{uw} > \nu R_u$. Then:
        \begin{enumerate}
            \item[$\bullet$] If $R_{\Delta} > 1$  and  $R_w \in I_1\cap I_2$, where $I_1 = (\nu R_u-R_{uw}+ 2\sqrt{\omega}, (2 - \nu)R_u -R_{uw} + 2\sqrt{\omega})$, $I_2 = (\nu R_u-R_{uw}, (2 - \nu)R_u -R_{uw} -2\sqrt{\omega})$ and $\omega=R_{uw} R_u (1 - \nu)(R_{\Delta} - 1)$, then there exist two positive coexistence  equilibria $P_{(u,w)}^+$ and $P_{(u,w)}^-$, with
            \[
            \bar{O}_{\pm} = \frac{k\lambda_i (R_w - 1)}{c} \left[ \frac{R_w + R_{uw} - \nu R_u \pm 2\sqrt{R_{uw} R_u (1 - \nu)(R_{\Delta} - 1)}}{2(1 - \nu)R_u} \right].
            \]
          \item[$\bullet$] If $R_{\Delta} = 1$ and $R_w\in\left(\nu R_u - R_{uw},(2 - \nu)R_u -R_{uw}\right)$, then there exists a unique positive coexistence equilibrium  with
            \[
            \bar{O} = \frac{k\lambda_i (R_w - 1)}{c} \left[ \frac{R_w + R_{uw} - \nu R_u}{2(1 - \nu)R_u} \right].
            \]
        \end{enumerate}
\item[2.2.] Suppose that $\underline{r} b \neq \underline{r}^w b^w$. Then:
        \begin{enumerate}
            \item[$\bullet$] If $R_{\Delta} > 1$, $S_1 < 0$, $S_2 < 0$, and $S_2 \mp S_3 \in (2S_1,\, 0),$ i.e., $S_2 \in (2S_1 - S_3 , - S_3) \cap (2S_1 + S_3 , 0)$, then there exist two positive equilibria $P_{(u,w)}^-$ and $P_{(u,w)}^+$ with:
            \[
            \bar{O}_{\pm} = \frac{k\lambda_i(R_w - 1)}{c} \left[ \frac{S_2 \mp S_3}{2S_1} \right].
            \]
            \item[$\bullet$] If $S_1 > 0 \Longrightarrow R_{\Delta} > 1$.  Therefore, if $S_2 \in (-S_3,\, 2S_1 - S_3)$, we have a unique positive coexistence equilibrium $P_{(u,w)}^-$     
with
            \[
            \bar{O}_- = \frac{k\lambda_i(R_w - 1)}{c} \left[ \frac{S_2 + S_3}{2S_1} \right].
            \]
            \item[$\bullet$] If $R_{\Delta} = 1$ and $S_2 \in (2S_1,0) \cup (0, 2S_1)$, then there exists a unique positive coexistence equilibrium with
            \[
            \bar{O} = \frac{k\lambda_i(R_w - 1)}{c} \frac{S_2}{2S_1}. 
            \]
        \end{enumerate}
    \end{enumerate} 
In any case, the other components can be obtained by substituting $\bar O$ into Eq.~\eqref{COEX}.

\section{Stability Analysis of the Equilibrium Points} 
Here, the local asymptotic stability of the equilibrium points $P_{(0,0)}$, $P_{(u,0)}$, and $P_{(0,w)}$ are obtained. \label{satibility_conditions}

\subsection{The Equilibrium Free of \emph{Wolbachia} Infection} 

Given the three dimension infected subsystem - $(O^w, I^w, A^w)$ - that describe the production of new
infections and changes in the state among infected individuals, we can use the next-generation matrix to obtain the asymptotic stability of $P_{(u,0)}$ \citep{van2002reproduction,diekmann1990definition}.
The subsystem is given by
\begin{eqnarray}
&&\frac{dO^w}{dt}= \phi^w \zeta r^wA^w - O^w\lambda_o^w\nonumber \\
&&\frac{dI^w}{dt}= \sigma_o^w O^w \left(1- \frac{I+I^w}{k}\right)- I^w\lambda_i^w \nonumber\\
&&\frac{dA^w}{dt}= \sigma_i^w I^w- \mu_a^wA^w,
\label{NGM}
 \end{eqnarray}
 and the matrices $F$ and $V$ are
 given by 
  \[
 F=\begin{pmatrix}
 0&0&\phi^w \zeta r^w\\
 0&0&0\\
 0&0&0
 \end{pmatrix} \quad \mbox{and}
 \quad V=\begin{pmatrix}
 \lambda_o^w&0&0\\
 -\sigma_o^w\left(1-\frac{I +  I^w}{k}\right)&\lambda_i^w&0\\
 0&-\sigma_i^w&\mu_a^w
 \end{pmatrix}.
 \]

 Remember that these two matrices are the decomposition of the Jacobian matrix in two parts: $F$ is the transmission and $V$ is the transition.  These two matrices evaluated at the disease-free equilibrium point $P_{(u,0)}=(\bar O, \bar I, \bar A, \bar Q, 0,0,0)$ are 
 \[
 F= \begin{pmatrix}
 0&0&\phi^w \zeta r^w\\
 0&0&0\\
 0&0 &0
 \end{pmatrix} \quad \mbox{and} \quad V= \begin{pmatrix}
\lambda_o^w&0&0\\
-\sigma_o^w\left(1-\frac{\bar I}{k}\right)&\lambda_i^w&0\\
 0&-\sigma_i^w&\mu_a^w 
 \end{pmatrix},
 \]
and the spectral radius of $FV^{-1}$ (dominant eigenvalue) gives
\[
\frac{R_w}{R_u}= \frac{\phi^w \zeta r^w \sigma_o^w\left(1-\frac{\bar I}{k}\right)\sigma_i^w}{\lambda_o^w\lambda_i^w\mu_a^w} \quad \mbox{with} \quad I=\frac{k(R_u-1)}{R_u}.
\]
Therefore, the existence and the local asymptotic stability of $P_{(u,0)}$ are guaranty by 
\begin{equation}
\label{R1}
R_u > R_w   \quad \mbox{and} \quad R_u>1.
\end{equation}

\subsection{The Trivial Equilibrium}

In this case, local asymptotic stability of $P_{(0,0)}$ is approached by the Jacobian matrix evaluated at this point and by the Routh–Hurwitz stability criterion for polynomials of degrees $3$ and $4$.  Because we have a singularity when $\bar A=\bar A^w=0$, let's suppose that $\bar A=\epsilon$, $\bar A^w=0$  and analysis what happened when $\epsilon \rightarrow 0.$  Firstly, note that
\[
\boldsymbol{J_0}=\left(
\begin{array} {cc}
\boldsymbol{M}& \boldsymbol{N} \\
\boldsymbol{0}&\boldsymbol{P}\\
\end{array}
\right)
\]
where
\[
\boldsymbol{M}=\left(
\begin{array} {cccc}
-\lambda_o& 0 & \phi r &0\\
\sigma_o&-\lambda_i&0& \sigma_q\\
0& \sigma_i&-\mu_a&0\\
\eta_o&0&0&-\lambda_q
\end{array}
\right), \quad \boldsymbol{N}=\left(
\begin{array} {ccc}
0& 0 & n_{13}\\
0&0&0\\
0&0&0\\
0&0&0\\
\end{array}
\right), \quad\mbox{and}\quad 
\boldsymbol{P}=\left(
\begin{array} {ccc}
-\lambda_o^w&0&\phi^wr^w \zeta\\
\sigma_o^w&-\lambda_i^w&0\\
0&\sigma_i^w&-\mu_a^w
\end{array}
\right).
\]with $n_{13}=\phi r \underline{r}^w(\nu -1)/\underline{r}+\phi^w\underline{r}^w\underline{\zeta}.$
Therefore, we have
\[
\det(\boldsymbol{J_0}-\lambda \boldsymbol{I})=\det(\boldsymbol{M}-\lambda \boldsymbol{I})\det(\boldsymbol{P}-\lambda \boldsymbol{I})=0,
\]
which give us 
\[
 \lambda^3+p_2 \lambda^2+p_1\lambda + p_0=0
\]
where
\begin{eqnarray}
&&p_2=\lambda_o^w+\lambda_i^w+\mu_a^w>0\nonumber\\
&&p_1=\lambda_o^w\lambda_i^w+\mu_a^w(\lambda_o^w+\lambda_i^w)>0\nonumber\\
&&p_0=\lambda_o^w\lambda_i^w\mu_a^w\left(1- R_w\right).\nonumber
\end{eqnarray}
and 
\[
 \lambda^4+m_3 \lambda^3+m_2 \lambda^2 + m_1 \lambda + m_0=0
\]
where
\begin{eqnarray}
m_3&=&\mu_a+\lambda_o+\lambda_i+\lambda_q>0\nonumber\\
m_2&=& (\lambda_q+\lambda_o)(\lambda_i+\mu_a)+\lambda_i\mu_a +\lambda_q\lambda_o>0\nonumber\\
m_1&=&\lambda_q[\lambda_o\lambda_i+\mu_a(\lambda_o+\lambda_i)]+\lambda_o\lambda_i\mu_a\left(1-\frac{\sigma_o}{c}R_u\right)\nonumber\\
m_0&=& \lambda_q\phi r \sigma_ic\left(\frac{1}{R_u}-1\right).\nonumber
\end{eqnarray}
If $ R_w <1$ then $p_0>0$ and  $p_1p_2>p_0$. Besides, if $R_u<1$, then $m_0>0$, $m_1>0$, $m_2m_3>m_1$, and  $m_1m_2m_3>m_1^2+m_0m_3^2$. Therefore, by the Routh-Hurwitz criterion, if $\max\{R_u, R_w \}<1$, then the trivial equilibrium is locally asymptotically stable.

Now, let's suppose that $\bar A=0$, $\bar A^w=\epsilon$  and analysis what happened when $\epsilon \rightarrow 0.$  We get
\[
\boldsymbol{\bar J_0}=\left(
\begin{array} {cc}
\boldsymbol{\bar M}& \boldsymbol{\bar N} \\
\boldsymbol{0}&\boldsymbol{\bar P}\\
\end{array}
\right)
\]
where
\[
\boldsymbol{\bar M}=\left(
\begin{array} {cccc}
-\lambda_o& 0 &\phi r \nu &0\\
\sigma_o&-\lambda_i&0& \sigma_q\\
0& \sigma_i&-\mu_a&0\\
\eta_o&0&0&-\lambda_q
\end{array}
\right), \quad \boldsymbol{\bar N}=\left(
\begin{array} {ccc}
0& 0 & \bar n_{13}\\
0&0&0\\
0&0&0\\
0&0&0\\
\end{array}
\right), \quad\mbox{and} \quad
\boldsymbol{\bar P}=\left(
\begin{array} {ccc}
-\lambda_o^w&0&\phi^wr^w \zeta\\
\sigma_o^w&-\lambda_i^w&0\\
0&\sigma_i^w&-\mu_a^w
\end{array}
\right).
\]with $\bar n_{13}=\phi^w r^w \underline{\zeta}$.

Therefore, we have
\[
\det(\boldsymbol{\bar J_0}-\lambda \boldsymbol{I})=\det(\boldsymbol{\bar M}-\lambda \boldsymbol{I})\det(\boldsymbol{\bar P}-\lambda \boldsymbol{I})=0,
\]
which give us 
\[
 \lambda^3+\bar p_2 \lambda^2+\bar p_1\lambda + \bar p_0=0
\]
where
\begin{eqnarray}
&&\bar p_2=\lambda_o^w+\lambda_i^w+\mu_a^w>0\nonumber\\
&&\bar p_1=\lambda_o^w\lambda_i^w+\mu_a^w(\lambda_o^w+\lambda_i^w)>0\nonumber\\
&&\bar p_0=\lambda_o^w\lambda_i^w\mu_a^w\left(1-R_w\right)\nonumber
\end{eqnarray}
and 
\[
 \lambda^4+\bar m_3 \lambda^3+\bar m_2 \lambda^2 + \bar m_1 \lambda + \bar m_0=0
\]
where
\begin{eqnarray}
\bar m_3&=&\mu_a+\lambda_o+\lambda_i+\lambda_q>0\nonumber\\
\bar m_2&=& (\lambda_q+\lambda_o)(\lambda_i+\mu_a)+\lambda_i\mu_a +\lambda_q\lambda_o>0\nonumber\\
\bar m_1&=&\lambda_q[\lambda_o\lambda_i+\mu_a(\lambda_o+\lambda_i)]+\lambda_o\lambda_i\mu_a\left(1-\frac{\sigma_o  \nu}{c}R_u\right)\nonumber\\
\bar m_0&=& \lambda_q \lambda_o\mu_a\lambda_i\left(1-\nu R_u)\right).\nonumber
\end{eqnarray}
If $R_w<1$ then $\bar p_0>0$ and  $\bar p_1 \bar p_2>\bar p_0$. Besides, if $\nu R_u<1$, then $\bar m_0>0$, $\bar m_1>0$, $\bar m_2 \bar m_3>\bar m_1$, and  $\bar m_1 \bar m_2 \bar m_3>\bar m_1^2+\bar m_0 \bar m_3^2$. Therefore, by the Routh-Hurwitz criterion, if $\max\{R_w,\nu R_u\}<1$, then the trivial equilibrium is locally asymptotically stable. Note that $\nu R_u<R_u<1.$

\subsection{The \emph{Wolbachia}-infected Equilibrium}

The local asymptotic stability of $P_{(0,w)}$ is approached by the Jacobian matrix evaluated at this point and by the Routh–Hurwitz stability criterion for polynomials of degrees 3 and 4. Firstly, note that
\[
\boldsymbol{J_w}=\left(
\begin{array} {cc}
\boldsymbol{S}& \boldsymbol{0} \\
\boldsymbol{T}&\boldsymbol{U}\\
\end{array}
\right)
\]
where
\[
\boldsymbol{S}=\left(
\begin{array} {cccc}
-\lambda_o& 0 & \phi r \nu &0\\
\frac{\sigma_o}{R_w}&-\lambda_i&0& \frac{\sigma_q}{R_w}\\
0& \sigma_i&-\mu_a&0\\
\eta_o&0&0&-\lambda_q
\end{array}
\right), \quad \boldsymbol{T}=\left(
\begin{array}{cccc}
0&0&0&0\\
0&t_{22}&0&0\\
0&0&0&0
\end{array}
\right),\quad  \boldsymbol{U}=\left(
\begin{array} {ccc}
-\lambda_o^w&0&\phi^w\zeta r^w\\
\frac{\sigma_o^w}{R_w}&-\lambda_i^wR_w&0\\
0&\sigma_i^w&-\mu_a^w
\end{array}
\right),
\]
with $t_{22}=-\lambda_i^w(R_w-1).$

Therefore, we have
\[
\det(\boldsymbol{J_w}-\lambda \boldsymbol{I})=\det(\boldsymbol{S}-\lambda \boldsymbol{I})\det(\boldsymbol{U}-\lambda \boldsymbol{I})=0,
\]
which give us 
\[
 \lambda^3+u_2 \lambda^2+u_1 \lambda + u_0=0
\]
where
\begin{eqnarray}
&&u_2=\mu_a^w+\lambda_o^w+\lambda_i^wR_w\nonumber\\
&&u_1=\mu_a^w(\lambda_o^w+\lambda_i^wR_w)+\lambda_o^w\lambda_i^wR_w\nonumber\\
&&u_0=\lambda_i^w\lambda_o^w\mu_a^w(R_w-1),\nonumber
\end{eqnarray}
and 
\[
 \lambda^4+s_3 \lambda^3+s_2\lambda^2 + s_1 \lambda + s_0=0
\]
where
\begin{eqnarray}
s_3&=&\mu_a+\lambda_o+\lambda_i+\lambda_q>0\nonumber\\
s_2&=&(\lambda_q+\lambda_o)(\lambda_i+\mu_a)+\lambda_i\mu_a +\lambda_q\lambda_o>0\nonumber\\
s_1&=&\lambda_o\lambda_i(\mu_a + \lambda_q) + \mu_a\lambda_q(\lambda_o  +\lambda_i)-z \sigma_o, \quad z:=\phi r\nu\sigma_i/R_w\nonumber\\
&=&\lambda_q[\lambda_o\lambda_i+\mu_a(\lambda_o+\lambda_i)]+cz\left(\frac{ R_w}{\nu R_u}-\frac{\lambda_q\sigma_o}{\lambda_q\sigma_o+\sigma_q\eta_o}\right),  \quad \mbox{ with } \nu\neq 0.\nonumber\\
s_0&=& \lambda_o\lambda_i\lambda_q\mu_a-\sigma_o \lambda_q z - \eta_o \sigma_q z=\left(\frac{ R_w}{ \nu R_u}-1\right)zc\lambda_q.\nonumber
\end{eqnarray}
Given that $\zeta=1$ and $R_w>1$, we have $ u_i>0$, for $i=0,1,2,$ and $u_1u_2>u_0$ which guarantees that all eigenvalues of matrix $\boldsymbol{U}$'s  characteristic polynomial are negative.
For the matrix $\boldsymbol{S}$, the coefficients of the characteristic equation have to satisfy: (i)  $s_i>0$, for $i=0,1,2,3$, (ii) $s_2s_3>s_1$, and (iii) $s_1s_2s_3>s_1^2+s_0s_3^2$. It is easy to see that if $ R_w> \nu R_u$ with $\nu \neq 0$ then $s_0>0$  and $s_1>0$. Moreover, in the case where $\nu=0$, we have $z=0$ which also implies $s_0>0$ and $s_1>0$.  Some algebraic manipulations are needed to prove that the same threshold holds for the other two conditions. Note that $ R_w> R_u$ implies $ R_w>\nu R_u$. 

\end{appendices}

\end{document}